\documentclass{article}
\makeatletter

\@addtoreset{equation}{section}
\makeatother

\usepackage{bm}

\usepackage{amsmath , amssymb}
\usepackage{amsthm}
\usepackage{slashed}

\allowdisplaybreaks[1]

\usepackage[normalem]{ulem}  % \sout{old text} for strikeout

\usepackage[dvipdfmx]{graphicx}

\usepackage{enumerate}

\usepackage{comment}

\usepackage{here}

\usepackage{mathcomp}

\usepackage{latexsym}

\usepackage{braket}

\usepackage[dvipsnames]{xcolor}

\usepackage[varg]{txfonts}
\usepackage{url}

\usepackage{indentfirst}

\usepackage{cite}

\usepackage{authblk}

\usepackage[%
    setpagesize=false,
    bookmarks=true,%
    bookmarksdepth=tocdepth,%
    bookmarksnumbered=true,%
    colorlinks=true,%
    linkcolor=BrickRed,
    pdftitle={},%
    pdfsubject={},%
    pdfauthor={},%
    pdfkeywords={}%
   ]{hyperref}

\title{Contribution of the Weinberg-type Operator to atomic and nuclear electric dipole moments}    
\author[1]{Naohiro Osamura\footnote{osamura.naohiro.j2@s.mail.nagoya-u.ac.jp}}
\affil[1]{Department of Physics, Nagoya University, Furo-cho Chikusa-ku, Nagoya, 464-8602 Japan}
\author[2]{Philipp Gubler}
\affil[2]{Advanced Science Research Center, Japan Atomic Energy Agency, Tokai, Ibaraki 319-1195, Japan}
\author[3,4]{Nodoka Yamanaka}
\affil[3]{Kobayashi-Maskawa Institute for the Origin of Particles and the Universe (KMI), Nagoya University, Nagoya, 464-8602, Japan.}
\affil[4]{Nishina Center for Accelerator-Based Science, RIKEN, Wako 351-0198, Japan}

\begin{document}

\maketitle

\begin{abstract}
    The contribution of the CP violating three-gluon Weinberg operator, $\frac{1}{3!} w f^{abc} \epsilon^{\nu \rho \alpha \beta} G^a_{\mu \nu} G^b_{\alpha \beta} G^{c \mu}_{\rho}$, to the atomic and nuclear EDMs is estimated using QCD sum rules. After calculating the transition matrix element between the pion and the vacuum through the Weinberg operator, we obtain the long-range CP-odd nuclear force by determining the isovector CP-odd pion-nucleon vertex, using chiral perturbation theory at NLO. The EDMs of $^{199}$Hg, $^{129}$Xe, $^{225}$Ra, $^2$H, and $^3$He are finally given including comprehensive uncertainty analysis. While the leading contribution of the $^{199}$Hg EDM is given by the intrinsic nucleon EDM, that of $^{129}$Xe atom may be dominated by the one-pion exchange CP-odd nuclear force generated by the Weinberg operator. From current experimental data of the $^{199}$Hg atomic EDM, we obtain an upper limit on the Weinberg operator magnitude of $\left|w \right| < 4 \times 10^{-10} {\rm GeV}^{-2}$ if we assume that it is the only source of CP violation at the scale $\mu =1$ TeV.
    
\end{abstract}

\newpage

\tableofcontents

\section{Introduction}
\label{Introduction}
    
    The matter dominant Universe is a cosmologically important phenomenon which cannot be explained by the standard model (SM). Indeed, a large CP violation is required to realize the matter dominance according to Sakharov \cite{Sakharov:1967dj}. However, the CP violation of the SM does not fulfill this criterion \cite{Farrar:1993hn,Huet:1994jb} and hence the experimental search for  new physics beyond the SM is actively pursued.

    The electric dipole moment (EDM) \cite{He:1990qa,Bernreuther:1990jx,khriplovichbook,Ginges:2003qt,Pospelov:2005pr,Fukuyama:2012np,Engel:2013lsa,Hewett:2012ns,Yamanaka:2014mda,Yamanaka:2017mef,Chupp:2017rkp} is a CP-violating observable sensitive to new physics which has been explored experimentally in various systems. The measurements of the EDMs in atomic systems are especially attracting attention, thanks to their high sensitivity \cite{Graner:2016ses,Bishof:2016uqx,Allmendinger:2019jrk}, which can for certain regions of the parameter space, be higher than the experimental constraints obtained by the LHC experiments. We also note their SM background is very small \cite{Kobayashi:1973fv,Seng:2014lea,Yamanaka:2015ncb,Yamaguchi:2020eub,Ema:2022yra}.
    
\begin{figure}[H]
        \centering
        \includegraphics[scale = 0.1]{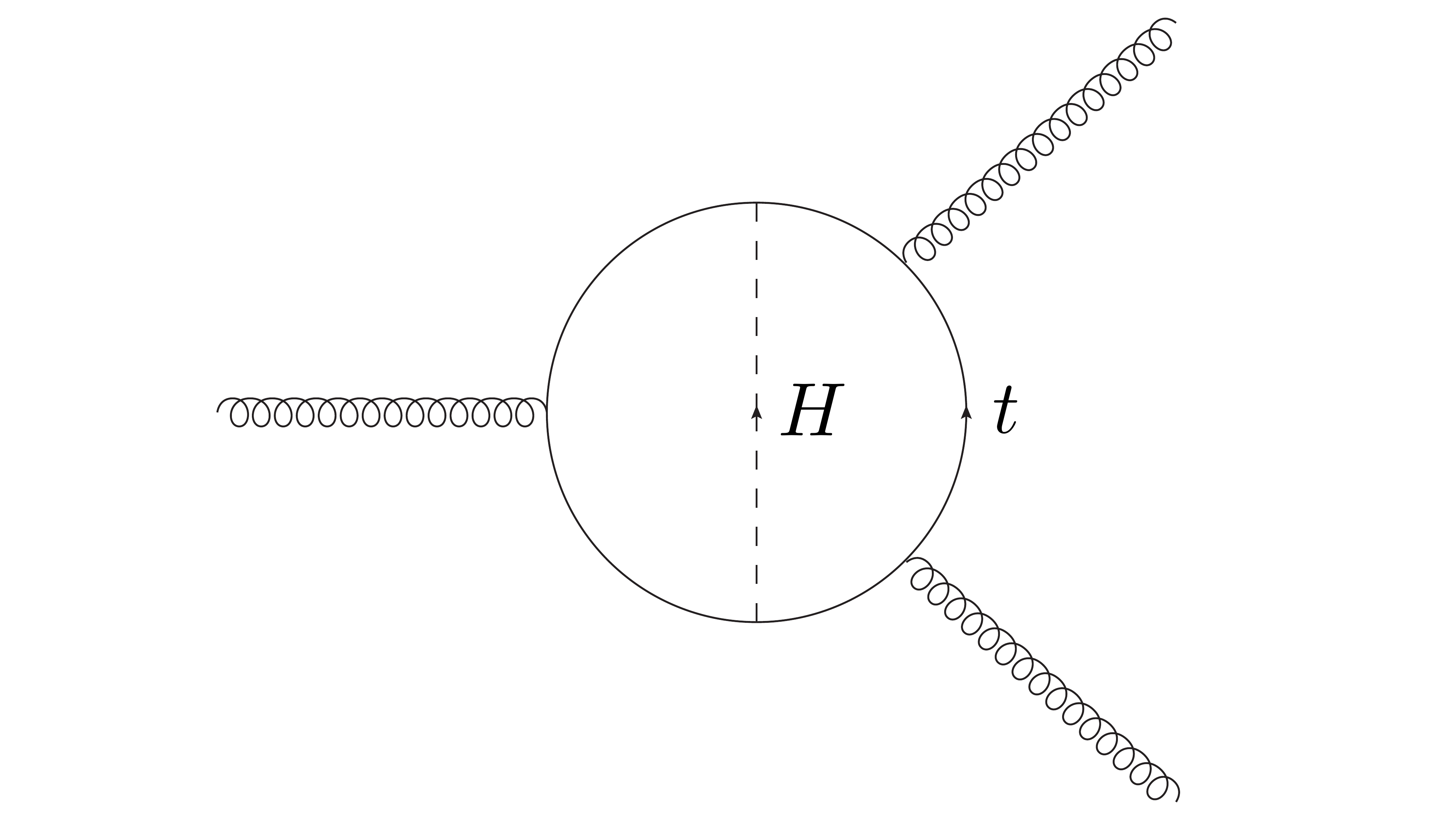}
        \caption{Two-loop level diagram contributing to the Weinberg operator generated in the two-Higgs doublet model.}
        \label{fig-two Higgs doublet}
    \end{figure}

    In practice one usually needs to integrate out new physics degrees of freedom to obtain an effective theory relevant at hadronic scales, which includes CP violating terms such as
    \begin{equation}
        \begin{split}
            \mathcal{L}_{\text{SMEFT,CP}} &= \frac{g_{s}^{2}}{32 \pi^{2}} \bar{\theta} G_{\mu \nu}^{a} \tilde{G}^{a \mu \nu} -\frac{i}{2} \sum_{i=u, d, s, e, \mu} d_{i} \bar{\psi}_{i} \sigma^{\mu} F_{\mu \nu} \gamma_{5} \psi_{i}-\frac{i}{2} \sum_{i=u, d, s} \tilde{d}_{i} \bar{\psi}_{i} g_{s} \sigma^{\mu \nu} G_{\mu \nu}^{a} \tau^{a} \gamma_{5} \psi_{i} \\
            &\quad +\frac{1}{3!} w f^{abc} \epsilon^{\nu \rho \alpha \beta} G^a_{\mu \nu} G^b_{\alpha \beta} G^{c \mu}_{\rho}+\sum_{i, j} C_{i j}\left(\bar{\psi}_{i} \psi_{i}\right)\left(\bar{\psi}_{j} i \gamma_{5} \psi_{j}\right) + \cdots.
        \end{split}
    \end{equation}
    Here, we include only low dimension terms which give the leading contribution to low energy observables. As for the diamagnetic atoms, the effect of the electron EDM is suppressed due to the closed electron shell, and the CP violation of the quark-gluon sector is important \cite{Ginges:2003qt,Yamanaka:2017mef,Fleig:2018bsf}. Contributions of the $\theta$ term \cite{Abramczyk:2017oxr,Dragos:2019oxn,Alexandrou:2020mds,Bhattacharya:2021lol} and the quark EDM \cite{Yamanaka:2018uud,Gupta:2018lvp,Alexandrou:2019brg,Cirigliano:2019jig,Horkel:2020hpi,Davoudi:2020ngi,Park:2021ypf} (second term in the above equation) to the nucleon EDM have already been extensively analyzed  in lattice QCD. The chromo-EDM (third term) is still difficult to handle on the lattice, but has been studied using QCD sum rules and chiral effective field theory ($\chi$EFT) \cite{Pospelov:2000bw,Pospelov:2001ys,Pospelov:2005pr,deVries:2010ah,Hisano:2012sc,deVries:2012ab,Fuyuto:2012yf,Mereghetti:2015rra,deVries:2015una,deVries:2015gea}. The purely gluonic CP-odd dimension-six Weinberg operator \cite{Weinberg:1989dx}, is less studied, but nevertheless important because it appears in many well-known models such as the Higgs doublet model (as shown in Fig.~\ref{fig-two Higgs doublet}) \cite{Weinberg:1989dx,Dicus:1989va,Boyd:1990bx,Cheng:1990gg,Bigi:1991rh,Bigi:1990kz,Hayashi:1994xf,Hayashi:1994ha,Wu:1994vx,Jung:2013hka,Brod:2013cka,Dekens:2014jka,Cirigliano:2016nyn,Cirigliano:2016njn,Panico:2017vlk,Cirigliano:2019vfc,Haisch:2019xyi,Cheung:2020ugr}, supersymmetric models \cite{Dine:1990pf,Dai:1990xh,Arnowitt:1990je,Abel:2001vy,Demir:2002gg,Demir:2003js,Degrassi:2005zd,Abel:2005er,Ellis:2008zy,Li:2010ax,Zhao:2013gqa,Sala:2013osa,Hisano:2015rna,Nakai:2016atk}, and other models \cite{Chang:1990sfa,Rothstein:1990vd,Xu:2009nt,Choi:2016hro,Abe:2017sam,Dekens:2018bci,DiLuzio:2020oah,Dekens:2021bro,Gisbert:2021htg}. It is also generated by the heavy quark sector CP violation via renormalization group evolution \cite{Braaten:1990gq,Chang:1990jv,Kamenik:2011dk,Gisbert:2019ftm,Yan:2020ocy,Haisch:2021hcg}. The SM contribution to the Weinberg operator
    \begin{equation}
        \mathcal{L}_W \equiv \frac{1}{3!} w f^{abc} \epsilon^{\nu \rho \alpha \beta} G^a_{\mu \nu} G^b_{\alpha \beta} G^{c \mu}_{\rho},
    \end{equation}
    is known to be unobservably small \cite{Booth:1992tv,Pospelov:1994uf,Yamaguchi:2020dsy}. While the effect of the Weinberg operator to the nucleon EDM has already been evaluated \cite{Chemtob:1991vv,Demir:2002gg,Dib:2006hk,Haisch:2019bml,Hatta:2020ltd,Yamanaka:2020kjo,Hatta:2020riw,Weiss:2021kpt}, its contribution to the CP-odd nuclear force, which is expected to be one of the leading effects to the atomic and nuclear EDMs, is less known.
    
\begin{figure}[H]
        \centering
        \includegraphics[scale = 0.1]{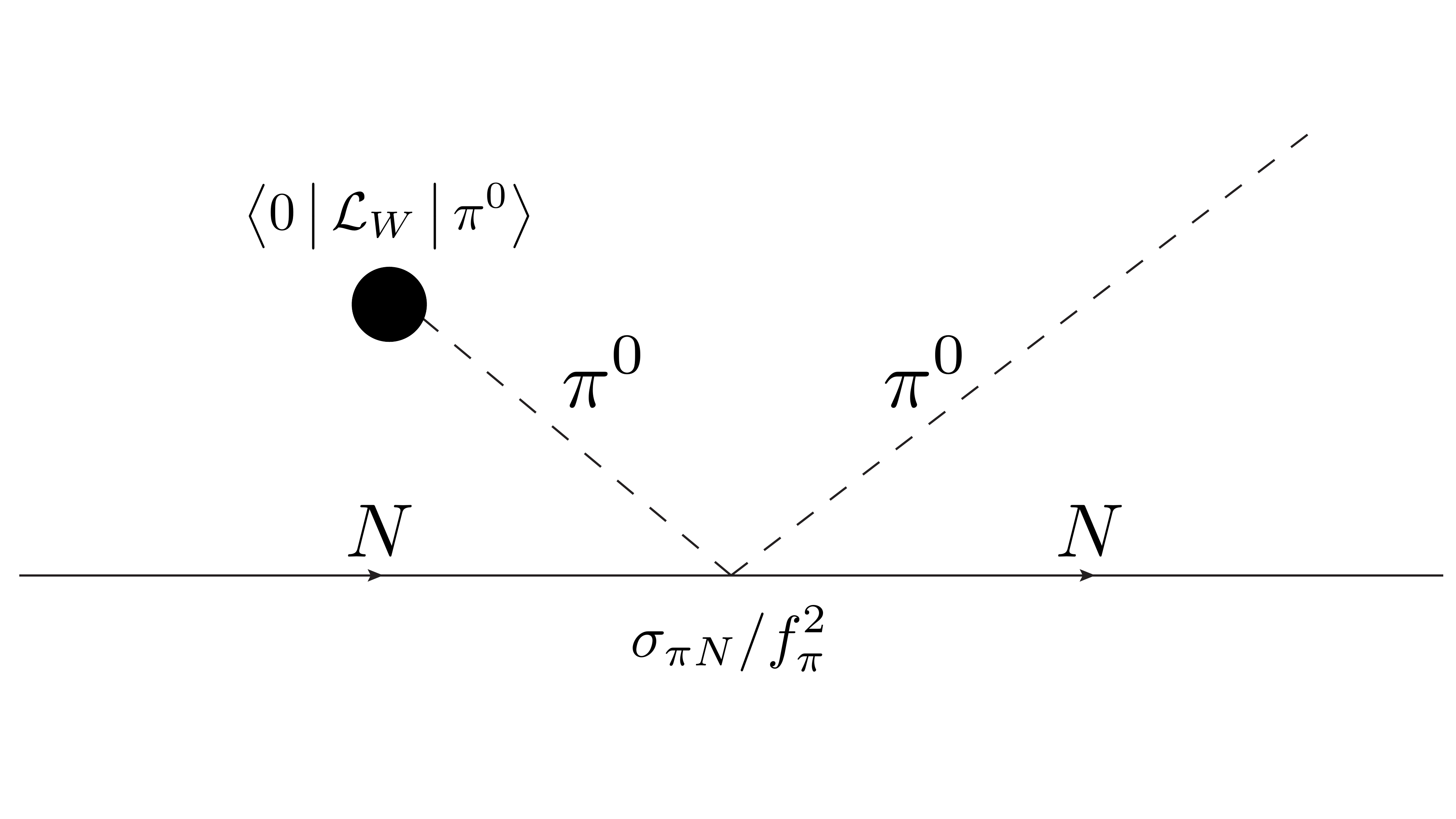}
        \caption{The leading order $\chi$EFT contribution to the isovector CP-odd $\pi N$ interaction through the Weinberg operator.}
        \label{fig-gbar}
    \end{figure}

    In $\chi$EFT, the leading contribution to the CP-odd nuclear force caused by the Weinberg operator is the contact nucleon-nucleon interaction \cite{deVries:2011an,Dekens:2014jka,Yamanaka:2016umw}, which is expected to have a small effect on the nuclear level CP violation due to the strong repulsion between nucleons. Similarly, the long range CP-odd nucleon-nucleon interaction generated by the pion exchange is also suppressed due to chiral symmetry. Indeed, the CP-odd pion-nucleon interaction, in contrast to the Weinberg operator, breaks chiral symmetry, so that the matching between them brings a suppression of at least one light quark mass factor. For these reasons, the contribution of the CP-odd nuclear forces has never been seriously considered in the context of atomic EDM, and it was long thought that the nucleon EDM is the leading process. However, a process generated by the pion-pole, as depicted in Fig.~\ref{fig-gbar}, is potentially significant because it may be enhanced by the large pion-nucleon sigma term. One also expects an enhancement of the CP-odd moments by the many-body effect for heavy atoms and nuclei.
    
    Let us briefly mention here that the quark EDM and chromo-EDM are also generated at low energy scales through the renormalization group running of the Weinberg operator even if the latter is the sole source of CP-violation at high energy \cite{Dine:1990pf,Degrassi:2005zd,Dekens:2013zca,Kley:2021yhn}. However, the quark EDM contribution is small at the hadron level \cite{Yamanaka:2018uud,Gupta:2018lvp,Alexandrou:2019brg,Cirigliano:2019jig,Horkel:2020hpi,Davoudi:2020ngi} and the calculation of the hadron matrix element of the chromo-EDM has a large theoretical uncertainty, some studies even predicting a null contribution \cite{Khatsimovsky:1987fr,Pospelov:2001ys,Pospelov:2005pr}.
    
    To investigate potentially significant effect mentioned above, we in this paper estimate the contribution of the Weinberg operator to the isovector CP-odd pion-nucleon interaction. To calculate the pion-vacuum transition matrix element $\Braket{\pi^0|\mathcal{L}_W|0}$, we employ QCD sum rules, which allows us to perform a relatively simple and analytic analysis  based on QCD \cite{Shifman:1978bx,Shifman:1978by,Gubler:2018ctz}. In the future, it might be possible to compute this matrix element from lattice QCD. However, this task is currently still difficult because of the computational cost of treating gluonic operators on the lattice and their accurate renormalization, although some first trials of such computation is already exist \cite{Cirigliano:2020msr,Rizik:2020naq}.
    \begin{figure}[H]
        \begin{minipage}[b]{0.45\linewidth}
            \centering
            \includegraphics[width = 0.78\linewidth]{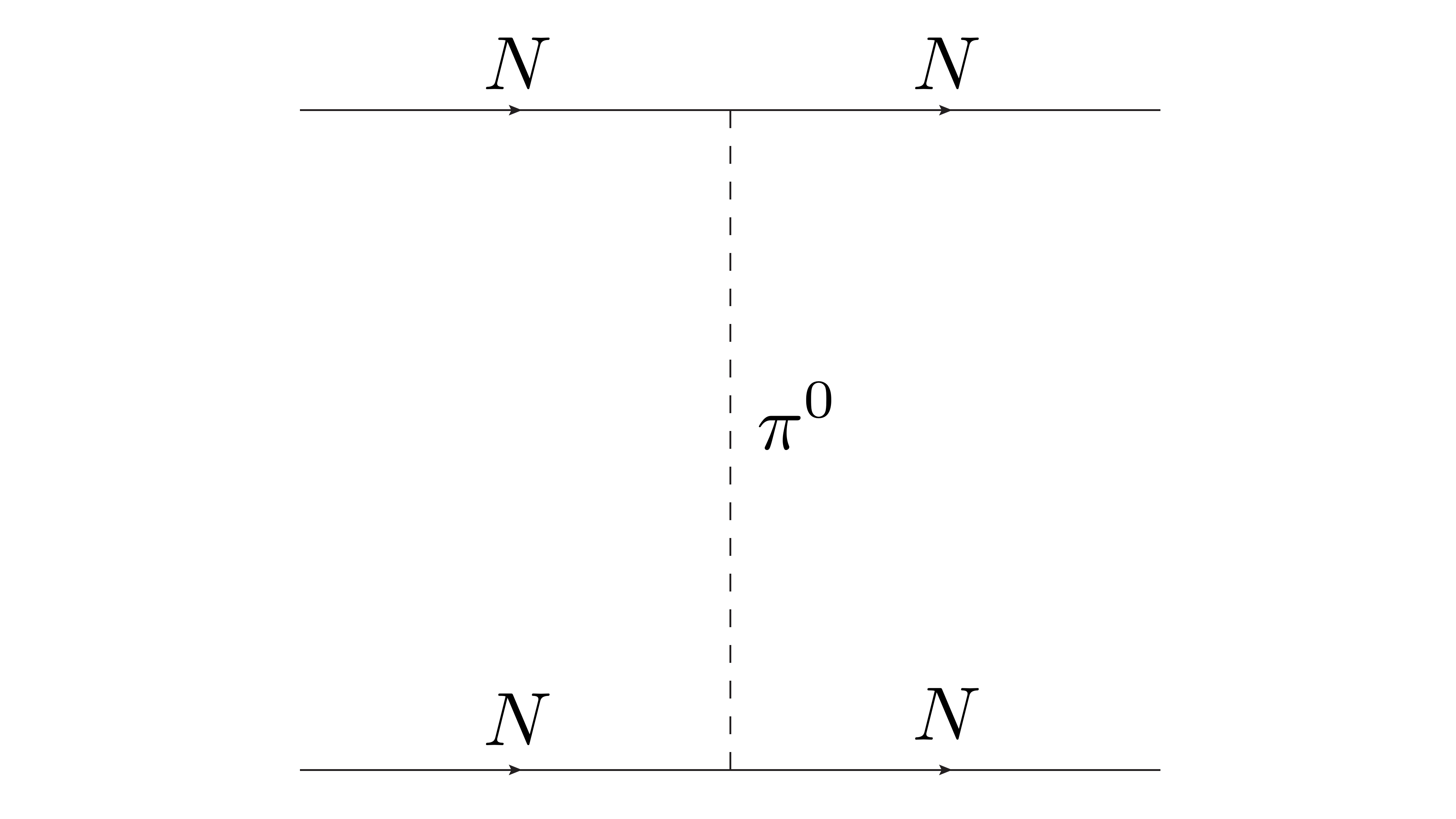}
            \label{fig-CPNNpi}
        \end{minipage}
        \begin{minipage}[b]{0.45\linewidth}
            \centering
            \includegraphics[width = 0.7\linewidth]{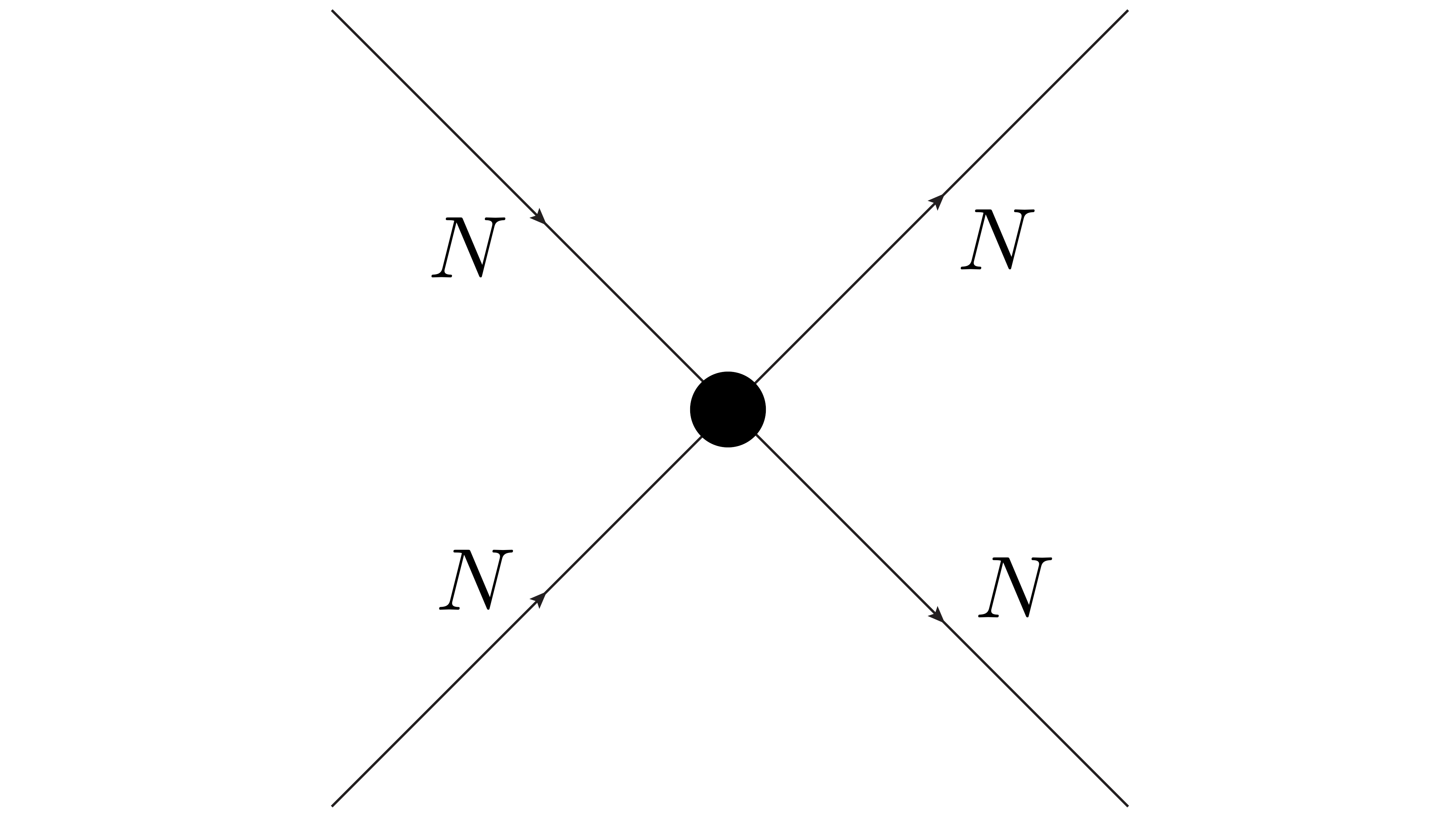}
            \label{fig-contactNN}
        \end{minipage}
        \caption{CP-odd nucleon-nucleon processes which induce nuclear level CP-odd moments. The left diagram is the long range nuclear force. It is generated by combining the CP-even and the CP-odd pion-nucleon interactions. The upper (lower) vertex represents the CP-even (CP-odd) interaction. The right figure depicts the CP-odd contact nucleon-nucleon interaction.}
    \end{figure}
    
    This paper is organized as follows. In Section~\ref{Formulation}, we set up the QCD sum rules formalism to compute the desired matrix element. The concrete calculation of the operator product expansion (OPE) and the numerical analysis follows in Sections~\ref{OPE calculation for Weinberg operators} and \ref{Numerical analysis}. In Section~\ref{Hadron level derivation of the one-pion exchange CP-odd nuclear force}, we derive the atomic and nuclear EDMs through hadron, nuclear and atomic level calculations, with a particular focus on the relative magnitudes between our newly calculated CP-odd nuclear force Weinberg operator contribution and that of the intrinsic nucleon EDM. The paper is concluded in Section~\ref{Conclusions}. Calculational details are provided in the Appendices.

\section{Formulation of the sum rules}
\label{Formulation}

     The goal of this and following two sections is the estimation of the Weinberg operator matrix element between the contribution of a pion state and the non-perturbative QCD vacuum $|0 \rangle$,
    \begin{equation}
        \Braket{\pi^0|\mathcal{L}_W|0}, 
        \label{al-matrix element}
    \end{equation}
    which, as mentioned in the introduction, contributes to atomic and nuclear EDMs through the isovector CP-odd pion-nucleon interaction.
    For this purpose, we will make use of the sum rule method. 
    At first sight, the most direct approach to extract this matrix element would be to start from a correlator of $\mathcal{L}_W$ and 
    the pionic operator $i\bar{q} \gamma_5 q$.
    However, such an estimation is challenging because it necessitates the computation 
    of a three-loop diagram at leading order in the OPE, which cannot be easily carried out using configuration space techniques \cite{Groote:2005ay}. We will therefore follow a different strategy extracting the pion pole from the correlation function of two Weinberg operators, 
    \begin{equation}
        \Pi (q^2) = i \int d^4 x e^{- i q \cdot x} \Braket{0|T \left[ \mathcal{L}_W (x) \mathcal{L}_W (0) \right]|0},
        \label{al-Correlation function of Weinberg}
    \end{equation} rather than computing the matrix element directly. $T$ here denotes the time ordering product.

\begin{figure}[H]
        \centering
        \includegraphics[width = 0.9\linewidth]{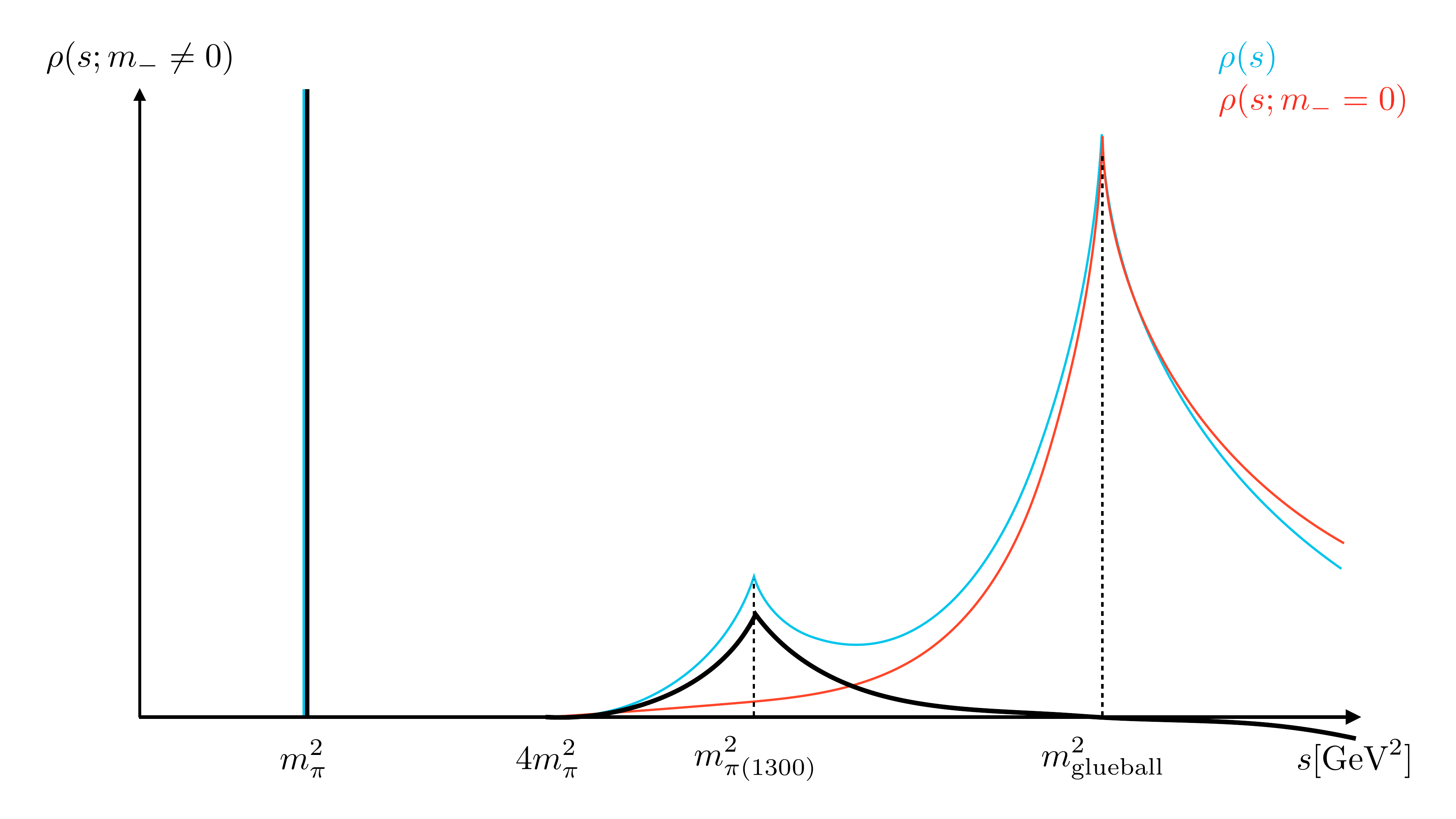}
        \caption{Schematic illustration of the isospin-breaking part of the spectral function $\rho (s;m_- \neq 0)$ (thick black line). The blue line depicts the complete spectral function $\rho(s)$ and red one indicates its isospin-symmetric part $\rho (s;m_-=0)$. $\pi(1300)$ is the first excited state of the pion. The $m_-^2$ differentiation causes only the isospin-breaking part survive.}
        \label{fig-spectrum function}
    \end{figure}

    We will estimate the correlator of Eq.~(\ref{al-Correlation function of Weinberg}) at large $-q^2$ (the deep-Euclidean region) using the OPE (see next section for details) and then 
    analyze the spectral function, defined as $\rho(s) = \frac{1}{\pi} \mathrm{Im} \Pi(s)$ numerically in Secion~\ref{Numerical analysis}. The dispersion relation is useful to relate the correlator in the deep-Euclidean region to the spectral function at $q^2 > 0$, where it 
    carries information about all physical states that couple to the operator $\mathcal{L}_W (x)$,
    \begin{equation}
        \begin{split}
            \Pi(q^2) &= \frac{1}{\pi} \int^\infty_0 ds \frac{\text{Im} \Pi (s)}{s-q^2} \\
            & = \int^\infty_0 ds \frac{\rho (s)}{s-q^2}.
            \label{al-dispertion relation}
        \end{split}
    \end{equation}
    As it is not possible to extract detailed features of the spectral function from Eq.~(\ref{al-dispertion relation}) and the (limited) OPE information available at large $-q^2$, we here approximate the spectral function to consist of a simple pion pole and a continuum structure above some effective threshold $s_{\text{th}}$ (for an alternative method making use of Bayesian statistics, see Ref.\cite{Gubler:2010cf}) . We furthermore assume that this continuum can be described by the OPE expression. We hence have
    \begin{equation}
        \begin{split}
            \rho_{\text{phen}} (s) &= \lambda_\pi^2 \delta (s - m_\pi^2) + \rho_{\text{cont}} (s) \theta (s-s_{\text{th}}) \\
            & = \lambda_\pi^2 \delta (s - m_\pi^2) + \frac{1}{\pi} \text{Im} \Pi_{\text{OPE}} (s) \theta (s-s_{\text{th}}),
            \label{phenomenological_side}
        \end{split}
    \end{equation}
    where $\lambda_\pi = \Braket{\pi^0|\mathcal{L}_W|0}$ is the quantity we seek to compute in this work. Note that the matrix element of Eq.~(\ref{al-matrix element}) must be proportional to $m_u - m_d \equiv m_-$ due to the chiral symmetry, and therefore is strongly suppressed. On the other hand, a pseudoscalar, glueball state can be expected to couple to the Weinberg operator and thus could 
    contribute significantly to the spectral function without any isospin suppression factor. Indeed, the correlator of Eq.~(\ref{al-Correlation function of Weinberg}) 
    is usually used to study the properties of such a glueball state (see, for example, Refs.\cite{Hao:2005hu,Chen:2021bck}). Hence, to suppress the potentially large glueball contribution of the sum rules, we will in this work make use of the smallness of the quark mass dependence of this state, which is confirmed in recent lattice QCD calculations \cite{Morningstar:1999rf,Chen:2005mg,Sun:2017ipk,Chen:2021dvn}. Specifically, we will study the quark mass dependence of the correlator by expanding it up to the second order in the up and down quark masses $m_u$ and $m_d$ and applying $m_-^2 \frac{\partial}{\partial m_-^2}$. This will strongly suppress the glueball contribution, while keeping the pion pole, for which $\lambda_\pi^2$ is proportional to $m_-^2$. More explicitly, in terms of the spectral function, $\rho (s)$ can be decomposed into two terms: $\rho (s) = \rho (s; m_-=0) + \rho (s; m_- \neq 0)$. The isospin-symmetric part $\rho (s;m_-=0)$ does not include the pion pole because of $\lambda_\pi = \Braket{\pi^0|\mathcal{L}_W|0} \propto m_-$. Thus, it at low energy only consists of a continuum spectrum beginning at the threshold $4 m_\pi^2$ and at higher energy of the glueball peak mentioned above. $\rho (s;m_-=0)$ furthermore will be positive definite since it is contains physical states in the limit of exact isospin symmetry. The isospin-breaking part $\rho (s;m_- \neq 0)$, illustrated as a thick black line in Fig.~\ref{fig-spectrum function}, can be obtained as $\rho (s) - \rho (s;m_-=0)$ and does in contrast not need to be positive definite. 
    As will be explicitly demonstrated later, the leading order (perturbative) OPE term leads to a negative high energy limit for $\rho (s;m_- \neq 0)$. 
    
    Finally, to both improve the convergence of the OPE and suppress the relative contributions of the higher energy states to the sum rules, we make use of the 
    Borel transform, which in this work is defined as
    \begin{equation}
        \mathcal{B} \left[ f(Q^2) \right] := \lim_{\substack{Q^2,n \to \infty \\ Q^2/n = M^2 = \text{const.}}} \frac{(Q^2)^n}{(n-1)!} \left( - \frac{d}{d Q^2} \right)^n f(Q^2), 
    \end{equation}
    where $Q^2 = -q^2$. After 
    substituting Eq.~(\ref{phenomenological_side}) into Eq.~(\ref{al-dispertion relation}), we obtain
    \begin{equation}
        \mathcal{B} \left[ \Pi_{\text{phen}} \right] (M^2) = \lambda_\pi^2 \cdot \frac{1}{M^2} e^{-m_\pi^2/M^2} + \frac{1}{\pi} \int^\infty_{s_{th}} ds \frac{e^{-s/M^2}}{M^2} \text{Im} \Pi_{\text{OPE}} (s;m_- \neq 0), 
    \end{equation}
    which gives the ``phenomenological side" of the sum rules. The computation of the ``theoretical side" will be discussed in the next section.

\section{OPE for the Weinberg operator correlator}
\label{OPE calculation for Weinberg operators}
    The OPE of the correlation function in Eq.~(\ref{al-Correlation function of Weinberg}) can formally be expressed as
    \begin{equation}
        \Pi_{\text{OPE}} (q) = i \int d^4 x e^{- i q \cdot x} \Braket{0|T \left[ \mathcal{L}_W (x) \mathcal{L}_W (0) \right]|0} = \sum_d C_d (q;\mu) \Braket{0|O_d (\mu)|0},
        \label{al-OPE}
    \end{equation}
    where $O_d$ stand for all local operators allowed by gauge invariance and vacuum symmetries while $C_d (q;\mu)$ represent their respective 
    Wilson coefficients. $\mu$ is the renormalization scale, which should be chosen large enough such that a perturbative calculation of the Wilson coefficient 
    is possible \cite{Novikov:1983jt}. Throughout this work, we will take $\mu = 1 {\rm GeV}$ if not explicitly stated otherwise.
    The Weinberg operator
    can for the purposes of the OPE calculation be rewritten as
    \begin{equation}
        \begin{split}
            \mathcal{L}_{W} & = \frac{1}{3!} w f^{abc} \epsilon^{\nu \rho \alpha \beta} G^a_{\mu \nu} G^b_{\alpha \beta} G^{c \mu}_{\rho} \\
            &= \frac{1}{3} w f^{abc} \epsilon^{\nu \rho \alpha \beta} g^{\mu \sigma} \left[ 2 \left( \partial_\mu A_\nu^a \right) \left( \partial_\rho A_\sigma^c \right) - \left( \partial_\mu A_\nu^a \right) \left( \partial_\sigma A_\rho^c \right) - \left( \partial_\nu A_\mu^a \right) \left( \partial_\rho A_\sigma^c \right) \right] \left( \partial_\alpha A_\beta^b \right) + \cdots.
        \end{split}
        \label{al-Weinberg operator2}
    \end{equation}
    $A_\mu^a$ here stand for gluon fields and the ellipses indicate higher order contributions with respect to the strong coupling constant.

\begin{figure}[H]
        \begin{minipage}[b]{0.24\linewidth}
            \centering
            \includegraphics[width=1\linewidth]{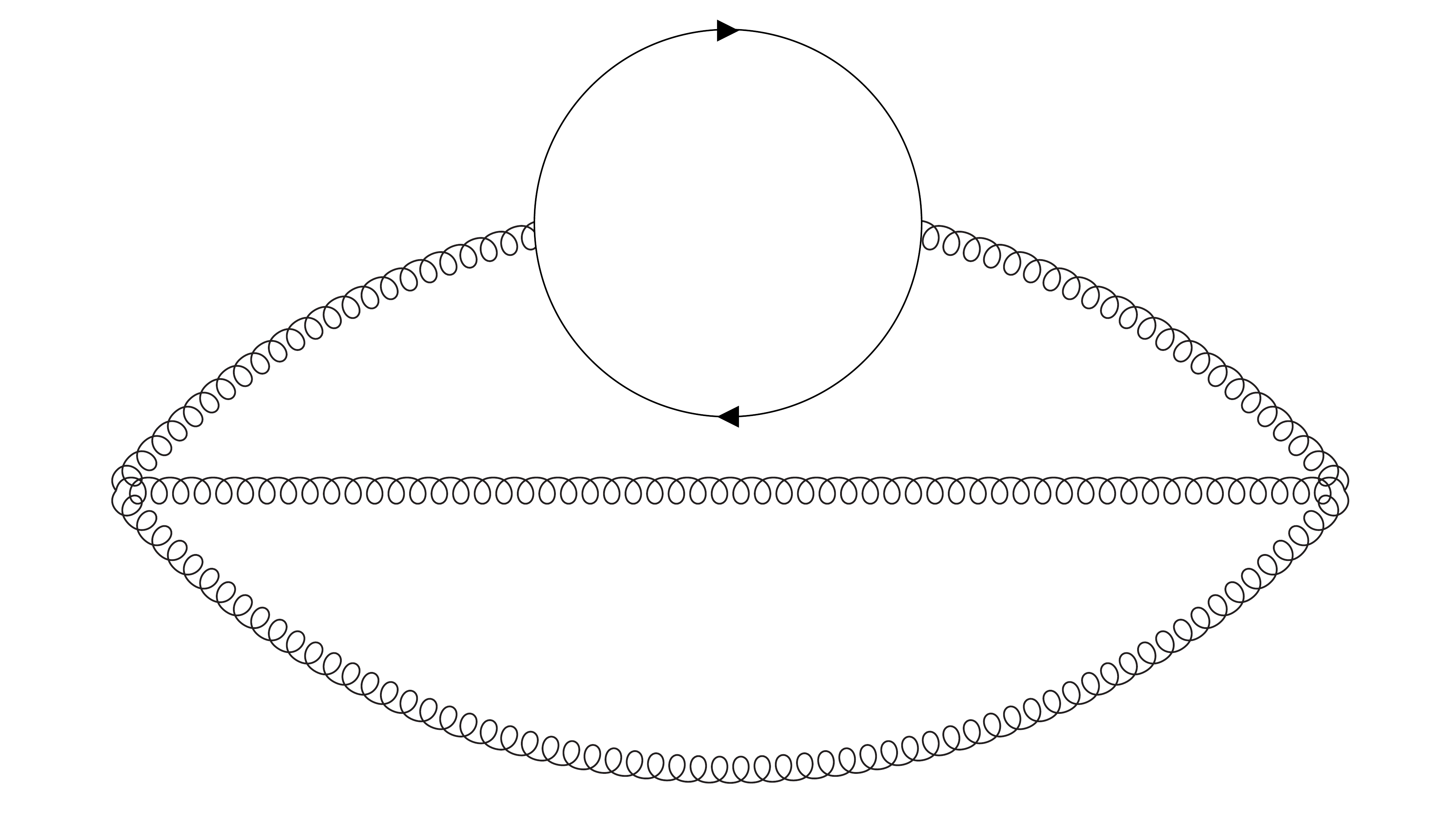}
            \label{fig-loop}
        \end{minipage}
        \begin{minipage}[b]{0.24\linewidth}
            \centering
            \includegraphics[width=1\linewidth]{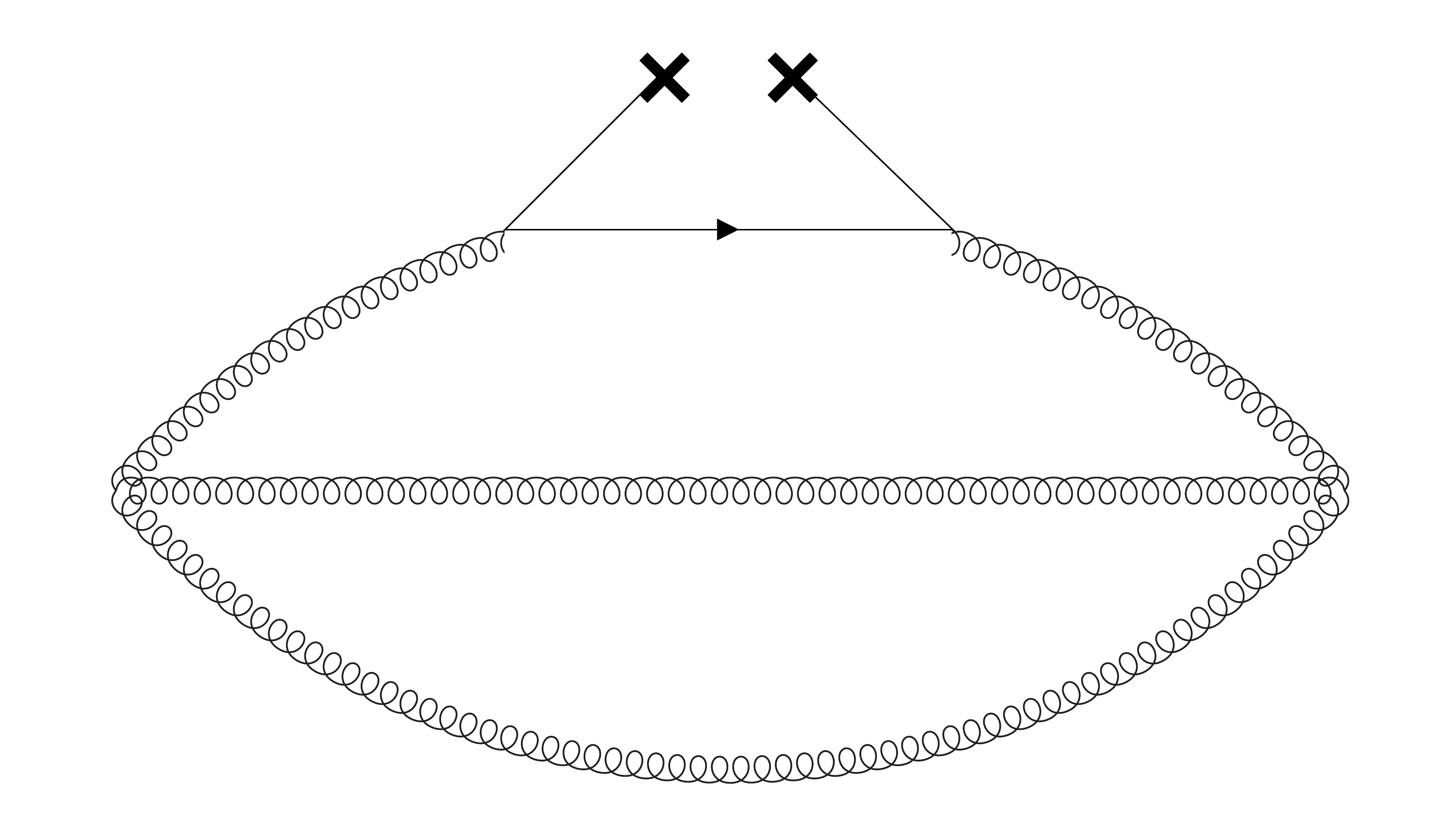}
            \label{fig-q}
        \end{minipage}
        \begin{minipage}[b]{0.24\linewidth}
            \centering
            \includegraphics[width=1\linewidth]{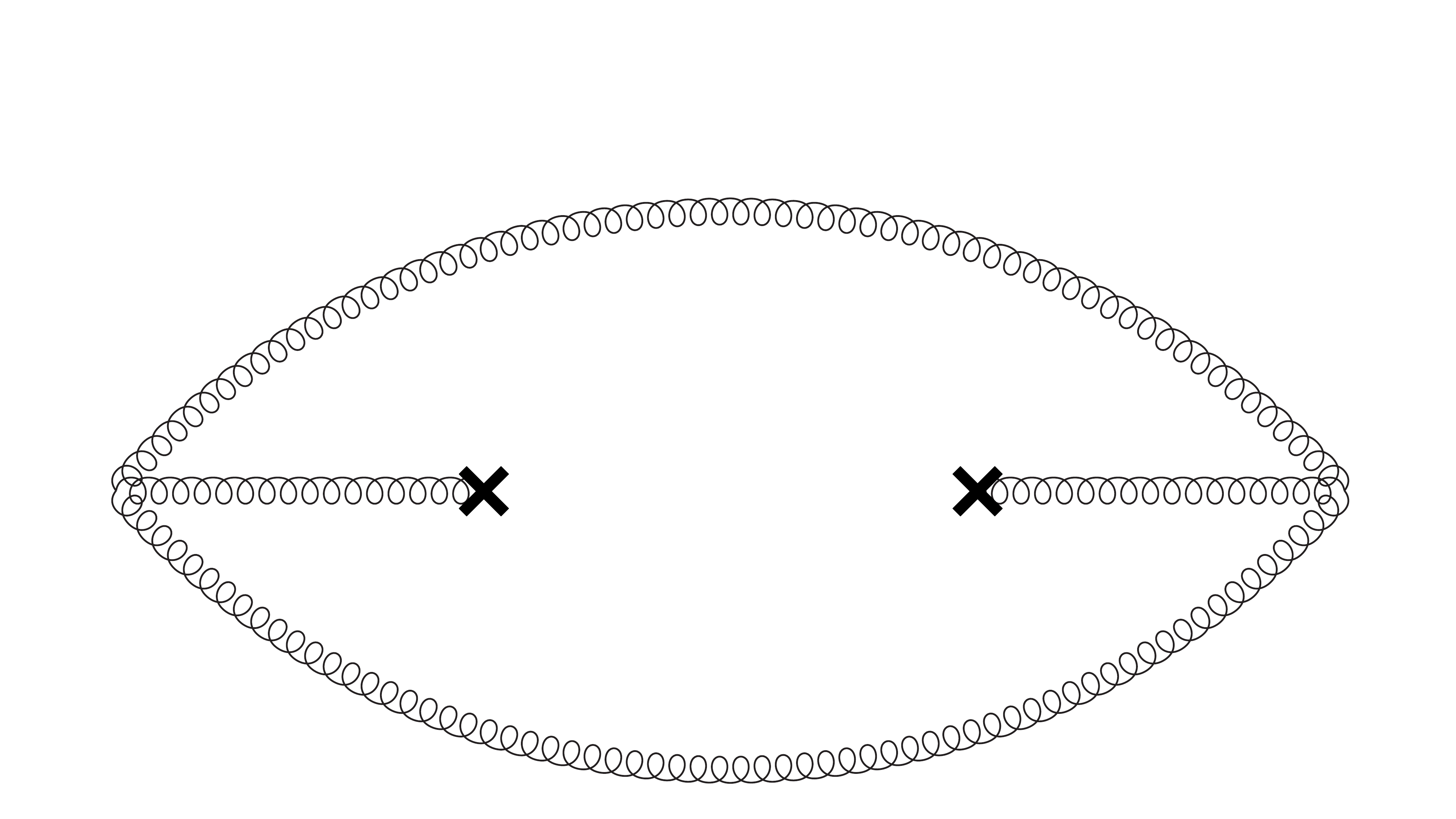}
            \label{fig-loop+G}
        \end{minipage}
        \begin{minipage}[b]{0.24\linewidth}
            \centering
            \includegraphics[width=1\linewidth]{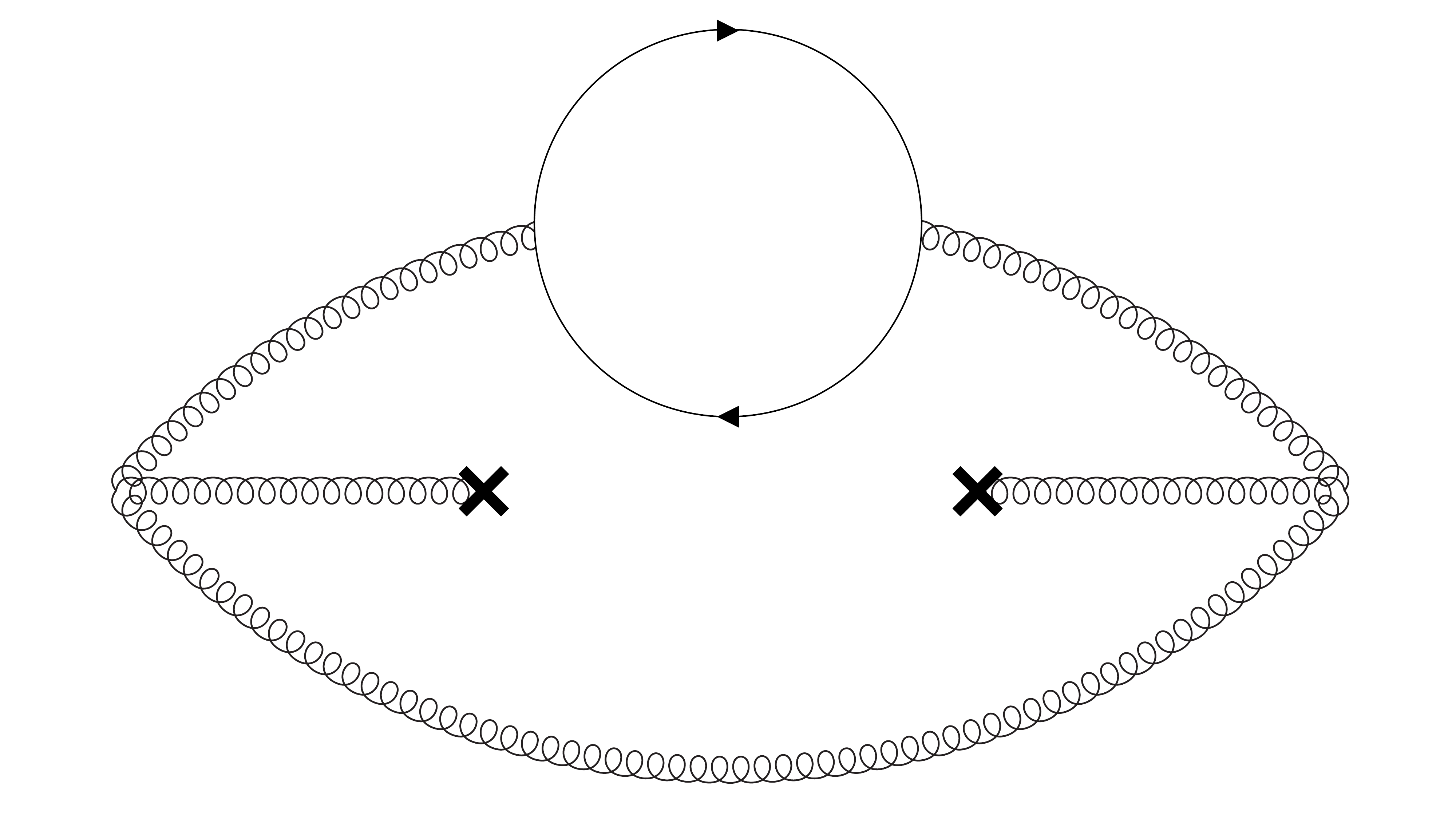}
            \label{fig-loop+G}
        \end{minipage}
        \caption{The diagrams corresponding to $\Pi_{\text{loop}} (q^2)$, $\Pi_{q} (q^2)$, $\Pi_{G} (q^2)$ and $\Pi_{\text{loop}+G} (q^2)$ of Eq.~(\ref{OPE_three_terms}), respectively. The crosses in the latter three diagrams represent the chiral condensate $\Braket{0|\bar{q}q|0}$, and the gluon condensate $\Braket{0|\alpha_s G^2|0}$. The two diagrams on the right, which both include a gluon condensate contribution, vanish in the OPE calculation of this work.}
        \label{fig-diagrams}
    \end{figure}

    As mentioned in the previous section, we are in this work only interested in the $m_-$-dependent of $\Pi(q^2)$ and hence decompose  $\Pi_{\text{OPE}}$ in the same way as the spectral function. Specifically, we have
    \begin{equation}
        \Pi_{\text{OPE}} (q^2) =\Pi_{\text{OPE}} (q^2;m_-=0) + \Pi_{\text{OPE}} (q^2;m_- \neq 0), 
    \end{equation}
    where the second term can be expanded using the OPE as
    \begin{equation}
        \Pi_{\text{OPE}} (q^2;m_- \neq 0) =  \Pi_{\text{loop}} (q^2) + \Pi_{q} (q^2) + \Pi_{G} (q^2) + \Pi_{\text{loop}+G} (q^2). 
        \label{OPE_three_terms}
    \end{equation}
    The four terms on the right-hand side correspond to the four diagrams shown in Fig.~\ref{fig-diagrams}.
    \noindent Essential parts of the computation of the four terms in  Eq.~(\ref{OPE_three_terms}) are provided  in Appendix~\ref{Details of OPE calculation}. The results read
    \begin{align}
        &\Pi_{\text{loop}} (q^2) = - w^2 \frac{ m_-^2 \alpha_s}{768 \pi^4} (-q^2)^3 \ln \left( -q^2 \right), \\
        &\Pi_{q} (q^2) = - w^2 \frac{m_-^2 B_0^2 h_3 \alpha_s}{4 \pi^3} (-q^2)^2 \ln \left( - q^2 \right), \label{quark_cond} \\
        &\Pi_{G} (q^2) = 0, \\
        &\Pi_{\text{loop}+G} (q^2) = 0.
    \end{align}
    where $B_0$ is defined by the relation $\frac{1}{2} \langle 0| \bar{u}u + \bar{d}d|0 \rangle = - B_0 f_\pi^2$. The parameter $h_3$ in Eq.~(\ref{quark_cond}) is related to the difference between the up and down quark chiral condensates (see Eqs.~(\ref{eq-qmassdependense},\ref{eq-parameters})). Specifically, we have \cite{Gasser:1983yg,Gasser:1984gg}
    \begin{equation}
        h_3 = - \frac{\Braket{0|\bar{u}u - \bar{d}d|0}}{4B_0^2(m_u - m_d)}.
    \end{equation}
    The $m_-$ dependent part of the correlator is thus obtained as
    \begin{equation}
        \begin{split}
            \Pi_{\text{OPE}} (q^2;m_- \neq 0) &= \Pi_{\text{loop}} (q^2) + \Pi_{q} (q^2) + \Pi_G (q^2) + \Pi_{\text{loop}+G} (q^2) \\
            & = w^2 \left[ - \frac{ m_-^2 \alpha_s}{128 \pi^4} (-q^2)^3 \ln \left( -q^2 \right) - \frac{m_-^2 B_0 h_3 \alpha_s}{4 \pi^3} (-q^2)^2 \ln \left( - q^2 \right) \right], 
        \end{split}
        \label{al-OPE correlation}
    \end{equation}
    which, after the Borel transform becomes
    \begin{equation}
        \begin{split}
            \mathcal{B} \left[ \Pi_{\text{OPE}} (q^2;m_- \neq 0) \right] = w^2 \left[ - \frac{ 3 m_-^2 \alpha_s}{64 \pi^4} M^6 + \frac{m_-^2 B_0^2 h_3 \alpha_s}{2 \pi^3} M^4 \right].
        \end{split}
        \label{al-OPE correlation borel tr}
    \end{equation}

\section{Numerical sum rule analysis}
\label{Numerical analysis}

        Having estimated both phenomenological and theoretical sides of the correlator of Eq.~(\ref{al-Correlation function of Weinberg}) in Sections~\ref{Formulation} and \ref{OPE calculation for Weinberg operators}, we can now proceed to the numerical analysis of the sum rules, which 
        after the Borel transform and the $m_-^2$ differentiation is obtained as
        \begin{equation}
            \frac{\lambda_\pi^{2}}{M^2} e^{-m_\pi^2/M^2} + \frac{1}{\pi} \int^\infty_{s_{th}} ds \frac{e^{-s/M^2}}{M^2} \text{Im} \Pi_{\text{OPE}} (s;m_- \neq 0) = \mathcal{B} \left[ \Pi_{\text{OPE}} (q^2;m_- \neq 0) \right] .
            \label{al-combine}
        \end{equation}
        From the above equation and Eqs.~(\ref{al-OPE correlation}) and (\ref{al-OPE correlation borel tr}), the pion pole residue $\lambda_\pi^2$ can be expressed as a function of the Borel mass $M$ and the threshold parameter $s_{\text{th}}$,
        \begin{equation}
  \begin{split}
      | \Braket{0|\mathcal{L}_{W}|\pi} |^2 &= \lambda_\pi^2  \\
      & = w^2 e^{m_\pi^2/M^2} \left[ \frac{ m_-^2 \alpha_s}{128 \pi^4} M^2 \left\{ e^{-\frac{s_{\text{th}}}{M^2}} \left(6 M^6+6 M^4 s_{\text{th}}+3 M^2 s_{\text{th}}^2+s_{\text{th}}^3\right) - 6 M^6 \right\} \right. \\
      & \left. - \frac{m_-^2 B_0^2 h_3 \alpha_s}{4 \pi^3} M^2 \left\{ e^{-\frac{s_{\text{th}}}{M^2}} \left(2 M^2 s_{\text{th}}+2 M^4+s_{\text{th}}^2\right) - 2 M^4 \right\}\right] .
    \end{split}
    \label{al-pion pole}
\end{equation}
        In the numerical analysis of Eq.~(\ref{al-pion pole}), we employ the following input parameters:
        \begin{table}[H]
            \centering
            \begin{tabular}{ccc}
                \hline
                Input parameters & Values & Ref. \\
                \hline
                $\alpha_s$ & $0.483 \pm 0.016$ & \cite{Bethke:2012jm,ParticleDataGroup:2020ssz} \\
                $- 4 (m_u -m_d) B_0 h_3/f_\pi^2$ & $[0.014,0.02]$ & \cite{GomezNicola:2010vgd} \\
                $m_\pi$ & $135$ MeV & \\
                $m_u$ & $2.9^{+0.66}_{-0.35}$ & \cite{Yamanaka:2016fjj} \\
                $m_d$ & $6.0^{+0.65}_{-0.23}$ & \cite{Yamanaka:2016fjj} \\
                $B_0 f_\pi^2$ & $(265 \text{MeV})^3$ & \cite{Yamanaka:2016fjj}
            \end{tabular}
            \caption{Input parameters used in the numerical sum rule analysis. All are given at the renormalization scale of $\mu = 1$ GeV.}
            \label{cond_value}
        \end{table}
        For the value range of the parameter $h_3$, we have used the $SU(2)$ estimate given in Ref.\,\cite{GomezNicola:2010vgd}. The quark masses 
        correspond to a renormalization scale of 1 GeV and were obtained by running their values from the 2 GeV value given in the PDG 
        to 1 GeV using a two-loop renormalization group equation \cite{Yamanaka:2016fjj}. $B_0$, which is related to the quark condensate, 
        similarly given at the renormalization scale of 1 GeV, 
        was obtained from the Gell-Mann-Oakes-Renner relation and a renormalization group rescaling \cite{Yamanaka:2016fjj}. 
        
\begin{figure}[H]
            \centering
            \includegraphics[width=0.8\linewidth]{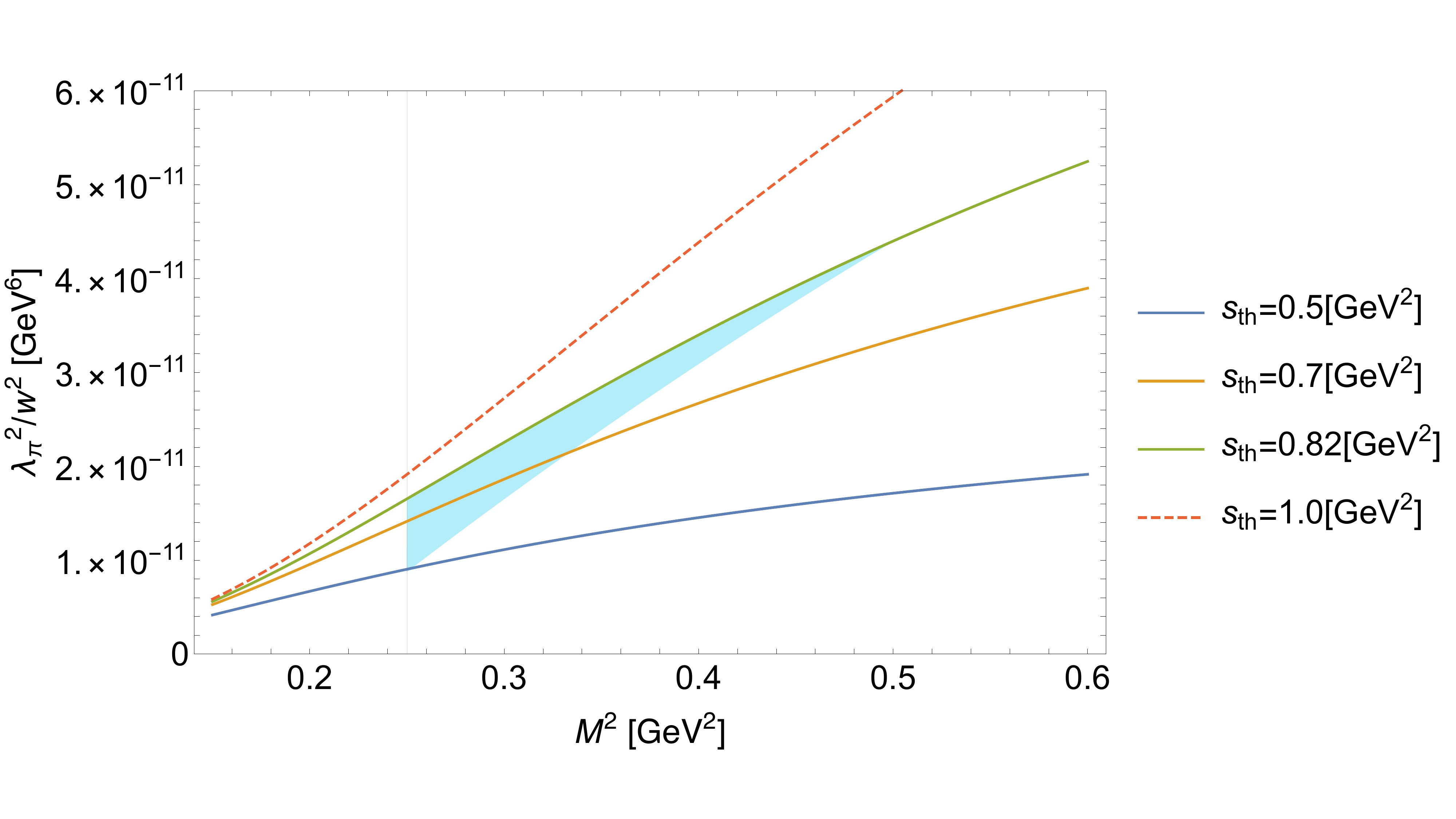}
            \caption{$\lambda_\pi^2/w^2$ as a function of the Borel mass $M$ for various values of the threshold parameter $s_{\text{th}}$. The blue shaded 
            region shows the parameter space of $M$ and $s_{\text{th}}$ used to estimate the range of $\lambda_\pi^2$.}
            \label{fig-pion pole}
        \end{figure}

        Let us here briefly discuss how to determine the parameter ranges of $M$ and $s_{\text{th}}$. For the 
        Borel mass $M$, one usually defines a so-called ``Borel window", within the various approximations used to derive 
        the sum rules are supposed to be valid.
        In our calculation, the term involving the quark condensate turns out to have the largest contribution. 
        As this term has the largest mass dimension in the final OPE result of Eq.~(\ref{al-OPE correlation borel tr}) and the gluon condensate terms vanish, it would 
        in this work not make much sense to determine the lower limit of the Borel window from the conventional condition of the OPE convergence. 
        We thus here consider only the upper limit, which can be fixed from the pole dominance criterion often used in the QCD sum rule literature, and set the lower limit by hand. The pole dominance criterion can be given as
        \begin{equation}
            \frac{\frac{\lambda_\pi^{2}}{M^2} e^{-m_\pi^2/M^2}}{\frac{\lambda_\pi^{2}}{M^2} e^{-m_\pi^2/M^2} + \frac{1}{\pi} \int^\infty_{s_{th}} ds \frac{e^{-s/M^2}}{M^2} \text{Im} \Pi_{\text{OPE}} (s;m_- \neq 0)} > 0.5, 
            \label{upper_limit_condition}
        \end{equation}
        which demands that the pion pole gives the dominant contribution to the phenomenological side of the sum rules.
        \begin{figure}[H]
            \centering
            \includegraphics[width = 0.8\linewidth]{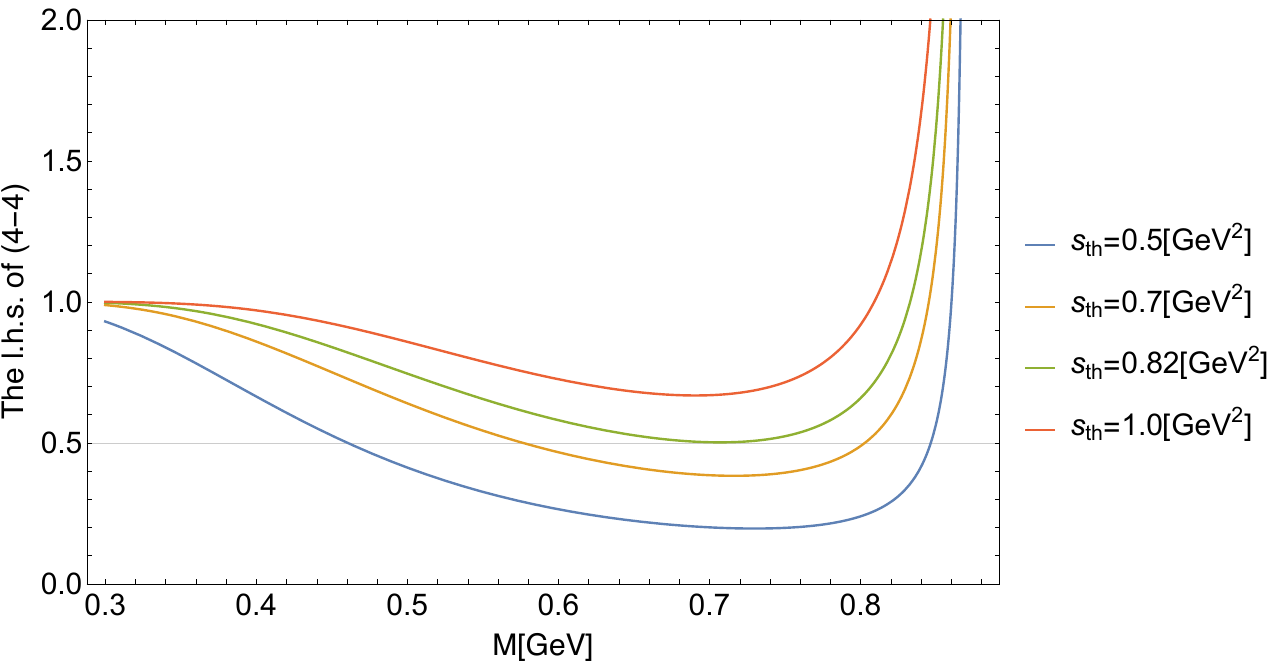}
            \caption{The left-hand side of Eq.~(\ref{upper_limit_condition}) as a function of the Borel mass $M$ for different values 
            of the threshold parameter $s_{\mathrm{th}}$.}
            \label{fig-upper limit}
        \end{figure}
        The left-hand side of Eq.~(\ref{upper_limit_condition}) is shown in Fig.~\ref{fig-upper limit}. We observe in this plot that the pole 
        contribution decreases with increasing Borel mass $M$ in the low $M$ region, which is a typical behavior in conventional QCD sum rule calculations. 
        For $M \gtrsim 0.8$ GeV, the ratio of Eq.~(\ref{upper_limit_condition}) 
        starts to increase because of a cancellation in its denominator and does not drop below 0.5 for too large values of the threshold parameter $s_{\mathrm{th}}$. We hence fix the 
        upper boundary of $s_{\mathrm{th}}$ such that this ratio can fall below 0.5, which gives $s^{\mathrm{max}}_{\mathrm{th}}=0.82\,\mathrm{GeV}^2$.
        The lower limit of $M$ is set up by hand as $M^2_{\text{min}}=0.25 \text{GeV}^2$. 
        The Borel window fixed in this way is depicted as the blue shaded region in Fig.~\ref{fig-pion pole}, where the colored curves show 
        the right-hand side of Eq.~(\ref{al-pion pole}) as functions of the Borel mass for different values of $s_{\mathrm{th}}$. 
        \noindent One can then see in this plot that the Borel 
        window is no longer open for threshold parameter values below $s_{\mathrm{th}}=0.5\,\mathrm{GeV}^2$, which thus fixes its lower boundary. All this then leads to the following range of values for the desired matrix element:
        \begin{equation}
            \left|\Braket{0|\mathcal{L}_W|\pi^0}\right| \in w \cdot [2.3,8.3] \times 10^{-6} \text{GeV}^5. 
        \end{equation}
        
\section{Derivation of the atomic and nuclear EDMs}
    \label{Hadron level derivation of the one-pion exchange CP-odd nuclear force}

        \begin{figure}[H]
            \centering
            \includegraphics[scale = 0.11]{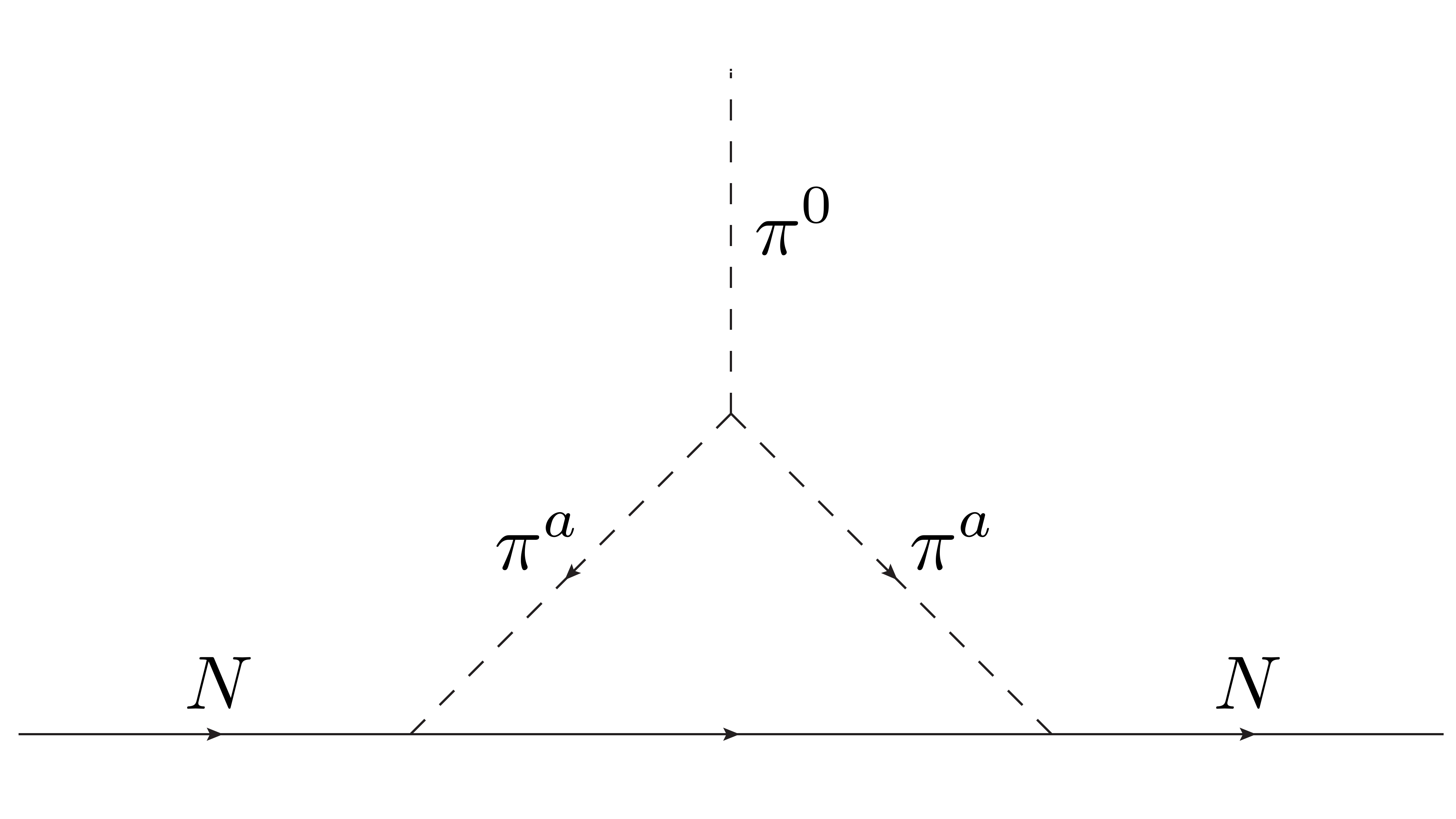}
            \caption{NLO correction to the isovector CP-odd pion-nucleon interaction. $\pi^a$ may be a charged or neutral pion.}
            \label{fig-gbar NLO}
        \end{figure}
    The CP-odd effective Lagrangian needed in this work is
    \begin{equation}
        \mathcal{L}_{CP} = -\frac{i}{2} d_{N} \bar{N} \sigma^{\mu} F_{\mu \nu} \gamma_{5} N+ k \, \pi^0+m_N \Delta_{3\pi}
    \pi_0(\pi_0^2 +2 \pi^+ \pi^-)  + \bar g_{\pi NN}^{(1)} \pi_0 \bar NN-\bar C_1 m_N \bar NN \, \bar Ni\gamma_5 N,
    \label{fig-LCP}
    \end{equation}
    where we implicitly assume that the sum over $N = n,p$ is taken. The first term of $\mathcal{L}_{CP}$ is the intrinsic nucleon EDM. The Weinberg operator contribution to the neutron EDM is given by\cite{Bigi:1991rh,Bigi:1990kz}
    \begin{equation}
        \begin{split}
            d_n &= \frac{m_{CP}}{m_N} \mu_n + d^{\text{(irr)}}_n \\
            & = w(20 \pm 12) e \, {\rm MeV}.
        \end{split}
        \label{eq:nucleonEDM}
    \end{equation}
    Here the first term corresponds to the chiral rotation of the neutron anomalous magnetic moment $\mu_n$ \cite{Demir:2002gg,Haisch:2019bml}, where $m_{CP}=-\left\langle N\left|\mathcal{L}_{w}\right| N\right\rangle=-m_{N} w \frac{3 g_{s} m_{0}^{2}}{32 \pi^{2}} \ln \left(\frac{M^{2}}{\mu_{\mathrm{IR}}^{2}}\right) = w \, (- 2.3 \pm 1.1 )\times 10^{-2} {\rm GeV}^3$ was calculated using QCD sum rules with $50 \%$ error according to Ref.~\cite{Haisch:2019xyi}. The second term stands for the irreducible term calculated in the quark model ($d^{\text{(irr)}}_n \approx -5\, w \, e$ MeV) \cite{Yamanaka:2020kjo}. We explicitly see from the above equation that the nucleon EDM generated by $w$ has no chiral suppression.
    
    The second term of $\mathcal{L}_{CP}$ is the neutral pion tadpole, which we have calculated in the previous section. Its coupling constant $k$ is the matrix element studied in this paper,
    \begin{equation}
    k
    =
    \langle 0 \,|\, 
    {\cal L}_W 
    \,|\, 
    \pi^0 \rangle
    ,
    \label{eq:pivacuum}
    \end{equation}
    which generates a neutral pion from the vacuum, and yields the isovector CP-odd pion-nucleon interaction (the term with $\bar{g}_{\pi NN}^{(1)}$) in combination with a pion-nucleon scattering process (see Fig.~\ref{fig-gbar}).
    The coupling of the pion-nucleon scattering is given by the matrix element
    \begin{equation}
    \langle \pi N
    \,|\, 
    {\cal L}_{\rm QCD} 
    \,|\, 
    \pi N \rangle
    \approx
    \frac{1}{f_\pi^2}
    \langle N
    \,|\, 
    m_u \bar u u 
    + m_d \bar d d 
    \,|\, 
    N \rangle
    \approx 
    \frac{\sigma_{\pi N}}{f_\pi^2}
    ,
    \label{eq:pi-N_scattering}
    \end{equation}
    where the first approximation is due to the partial conservation of the vector current, with $f_\pi =93$ MeV.
    The low energy constant $\sigma_{\pi N}$ appearing on the right-hand side of the above equation is the pion-nucleon sigma term, and represents the contribution of the current quark mass to the nucleon mass.
    Its value is still under debate, phenomenological studies yielding $\sigma_{\pi N} \approx 60$ MeV \cite{Hoferichter:2015dsa,Friedman:2019zhc} (see however Ref.~\cite{Huang:2019not} for a smaller prediction), whereas lattice QCD results point to a smaller value $\sigma_{\pi N} \approx 30$ MeV \cite{Yang:2015uis,Yamanaka:2018uud,Alexandrou:2019brg}.
    A recent lattice QCD analysis is suggesting that this discrepancy may come from the contamination of excited states with additional pions of the correlators computed on the lattice \cite{Gupta:2021ahb}.
    Combining the neutral pion generated by Eq.~(\ref{eq:pivacuum}) and the pion-nucleon scattering (\ref{eq:pi-N_scattering}), we obtain the isovector CP-odd pion-nucleon coupling as
    \begin{equation}
    \bar g_{\pi NN}^{(1)}
    =
    \langle 0 \,|\, 
    {\cal L}_W 
    \,|\, 
    \pi^0 \rangle
    \frac{\sigma_{\pi N}}{f_\pi^2 m_\pi^2}
    .
    \label{eq:piNN}
    \end{equation}

    To estimate the theoretical uncertainty of the above LO $\chi$EFT, we will in the following evaluate the NLO contribution to $\bar g_{\pi NN}^{(1)}$, which
    is in fact generated by the one-loop diagram given in Fig.~\ref{fig-gbar NLO}, via the three-pion interaction (the term with $\Delta_{3\pi}$ in Eq.~(\ref{fig-LCP})) \cite{deVries:2012ab,Yamanaka:2016umw}, which
    is part of the linear chiral symmetry and CP breaking term of the chiral Lagrangian
    \begin{equation}
    {\cal L}_{2CP}
    =
    {\rm Tr} [U A -U^\dagger A]
    =
    \frac{4 ia}{f_\pi}\pi_0
    -\frac{4 ia}{3! f_\pi^3} \pi_0(\pi_0^2 +2 \pi^+ \pi^-) 
    +\cdots
    ,
    \label{eq:L2CP}
    \end{equation}
    where $U\equiv e^{i\sqrt{2}M/f_\pi}$, with the meson matrix
    \begin{equation}
    M = 
    \left(
    \begin{array}{cc}
    \pi_0/\sqrt{2} &  \pi^+ \cr
    \pi^- & -\pi_0/\sqrt{2}  \cr
    \end{array}
    \right)
    .
    \end{equation}
    Here we only considered the $SU(2)_L \times SU(2)_R$ subspace for simplicity.
    The matrix $A = a \tau_z$ expresses the explicit isospin breaking generated by the Weinberg operator. By matching Eq. (\ref{eq:L2CP}) and the pion one-point interaction (\ref{eq:pivacuum}), we obtain 
    \begin{equation}
    a = 
    \frac{f_\pi}{4i}
    \langle 0 | {\cal L}_W | \pi_0 \rangle
    .
    \end{equation}
    The three-pion coupling generated by the Weinberg operator is then
    \begin{equation}
    m_N \Delta_{3\pi}
    = 
    -\frac{\langle 0 | {\cal L}_w | \pi_0 \rangle}{6f_\pi^2}
    .
    \end{equation}
    
    Evaluating the zero-momentum transfer limit of the one-loop diagram of Fig.~\ref{fig-gbar NLO}, we obtain the NLO contribution to the isovector CP-odd pion-nucleon coupling as \cite{Bsaisou:2014zwa,Bsaisou:2014oka}
\begin{equation}
\bar g_{\pi NN,{\rm NLO}}^{(1)}
=
-
m_N \Delta_{3\pi}
\frac{15 g_A^2 m_\pi}{32 \pi f_\pi^2}
=
+
\frac{5 g_A^2 m_\pi}{64 \pi f_\pi^4}
\langle 0 | {\cal L}_w | \pi_0 \rangle
,
\label{eq:piNNNLO}
\end{equation}
where $g_A = 1.27$.
The magnitude of the above NLO result amounts to about 20\% of the LO expression of Eq.~(\ref{eq:piNN}) (if we assume $\sigma_{\pi N}= 60$ MeV). This relatively large one-loop correction is due to the factor $4\pi$ enhancement arising from the heavy baryon approximation.
Regarding the sign of the Eq.~(\ref{eq:piNNNLO}), this one-loop level contribution is constructive, in accordance with Refs. \cite{Bsaisou:2014zwa,Bsaisou:2014oka} (for details, see Appendix~\ref{eq:NLOgpiNN}).

    The CP-odd pion-nucleon interaction (\ref{eq:piNN}) and (\ref{eq:piNNNLO}) generates an isovector type CP-odd nuclear force through the one-pion exchange.
    Its Hamiltonian is given by
    \begin{equation}
    {\cal H}
    =
    \frac{\bar G_\pi^{(1)}}{8 \pi m_N}
    (\vec{\sigma}_1 \tau_{1z} - \vec{\sigma}_2 \tau_{2z})
    \cdot 
    \vec{\nabla } \frac{e^{-m_\pi r}}{r} 
    =
    \frac{-\bar g_{\pi NN}^{(1)} g_A}{8 \pi f_\pi}
    (\vec{\sigma}_1 \tau_{1z} - \vec{\sigma}_2 \tau_{2z})
    \cdot 
    \vec{\nabla } \frac{e^{-m_\pi r}}{r} 
    ,
    \label{eq:CPVNN}
    \end{equation}
    where $\vec{\sigma}_i$ and $\tau_{iz}$ ($i=1,2$) are the spin and isospin operators acting on the $i$-th interacting nucleon, respectively.
    The coordinate $\vec r$ is directed to nucleon 1, and $\vec{\nabla }$ is the gradient defined accordingly.
    Here we defined $\bar G_\pi^{(1)} \equiv -\frac{ g_A m_N}{f_\pi}\bar g_{\pi NN}^{(1)}$ as commonly done in nuclear level calculations.

    The CP-odd nuclear force of Eq. (\ref{eq:CPVNN}) polarizes the nucleus and leads to an observable effect.
    As for atomic systems, the EDMs of atomic nuclei are not directly observable due to the notorious Schiff's screening phenomenon \cite{Schiff:1963zz}. The residual CP-odd moment, the nuclear Schiff moment $\vec{S}^{\text{A}}$ (NSM), has to be evaluated using nuclear structure calculations.
    This has been done using several methods \cite{Ginges:2003qt,Dmitriev:2004fk,Dobaczewski:2005hz,Ban:2010ea,Yoshinaga:2013rva,Yamanaka:2017mef,Yanase:2020agg,Yanase:2020oos,Yanaseprivate}.
    Here we quote the most recent results for $^{199}$Hg \cite{Yanase:2020agg,Yanase:2020oos}, $^{129}$Xe \cite{Yanase:2020agg,Yanaseprivate}, and $^{225}$Ra \cite{Dobaczewski:2005hz} nuclei:
    \begin{eqnarray}
    S^{\rm Hg}
    &=&
    2.65\, d_n \, {\rm fm}^2
    - 0.075 \, \bar G_\pi^{(1)} e \, {\rm fm}^3
    ,
    \\
    S^{\rm Xe}
    &=&
    0.42 \, d_n \, {\rm fm}^2
    + 0.041 \, \bar G_\pi^{(1)} e \, {\rm fm}^3
    ,
    \\
    S^{\rm Ra}
    &=&
    -6.0 \, \bar G_\pi^{(1)} e \, {\rm fm}^3
    ,
    \end{eqnarray}
    where we only display the contributions from the isovector CP-odd nuclear force and from the intrinsic nucleon EDM, for which we estimate the relative error to be about 30\%.
    The contribution of the NSM to the atomic EDM has also been calculated within several frameworks.
    We here quote the results of the latest calculations \cite{Sakurai:2019vjs,Prasannaa:2020cjx,Hubert:2022pnl},
    \begin{align}
        & \begin{aligned}
            d^{\rm Hg}
            & = -2.4 \times 10^{-17} \, \frac{S^{\rm Hg}}{e\, {\rm fm}^3} e\, {\rm cm}
            = -6.4 \times 10^{-4} d_n +1.8 \times 10^{-5} \bar G_\pi^{(1)} \, e\, {\rm fm} \\
            & = w \left(- 1.3 \times 10^{-2} \pm [0.71,4.4] \times 10^{-3}  \right) e \text{ MeV},
        \end{aligned} \\
        & \begin{aligned}
            d^{\rm Xe}
            & = 0.32 \times 10^{-17} \, \frac{S^{\rm Xe}}{e\, {\rm fm}^3} e\, {\rm cm}
            = 1.3 \times 10^{-5} d_n +1.3 \times 10^{-6} \bar G_\pi^{(1)} \, e\, {\rm fm} \\
            & = w \left( 2.7 \times 10^{-4} \pm [0.52,3.2] \times 10^{-4} \right) e \, {\rm MeV},
        \end{aligned}\\
        & \begin{aligned}
            d^{\rm Ra}
            & = -6.3 \times 10^{-17} \, \frac{S^{\rm Ra}}{e\, {\rm fm}^3} e\, {\rm cm}
            = 3.8 \times 10^{-3} \bar G_\pi^{(1)} \, e\, {\rm fm} \\
            & = w \left( \pm [1.5,9.3] \times 10^{-1} \right) e \, {\rm MeV}.
        \end{aligned}
    \end{align}
    The atomic level calculations are in relatively good agreement among each other \cite{Ginges:2003qt,Dzuba:2009kn,Radziute:2013sba,Yamanaka:2017mef,Sahoo:2018ile}. Their error is therefore negligible in the error budget.

    It is also interesting to inspect the EDM of light nuclei for which experimental measurements are currently planned \cite{Farley:2003wt,Anastassopoulos:2015ura,JEDI:2016swi}.
    For the deuteron and the $^3$He nucleus, we have \cite{Yamanaka:2016umw}
    \begin{align}
        & \begin{aligned}
            d^{^2{\rm H}}
            &=
            0.014 \, \bar G_\pi^{(1)} \, e\, {\rm fm} \\
            &=w \left(
            \pm [0.56,3.4] \right) e \, {\rm MeV},
            \label{eq-mercury}
        \end{aligned}\\
        & \begin{aligned}
            d^{^3{\rm He}}
            &= 0.88 \, d_n
            +0.010 \, \bar G_\pi^{(1)} +[1,5] \times 10^{-4} \bar C_1 \, {\rm GeV}^3 \, e\, {\rm fm} \\
            &= w \left(18
            \pm [0.38,2.3] + [1,10]\right) e\, {\rm MeV}.
        \end{aligned}
    \end{align}
    The nuclear level calculations for these light nuclei were performed by many groups \cite{Song:2012yh,Bsaisou:2014zwa,Yamanaka:2015qfa,Gnech:2019dod,deVries:2020iea,Yang:2020ges,Froese:2021civ} and all are consistent, having errors of less than 10\%. There are further results for other nuclei which may have larger sensitivities to CP violation, but are not considered here \cite{Yamanaka:2015qfa,Yamanaka:2016itb,Yamanaka:2016umw,Lee:2018flm,Yamanaka:2019vec,Froese:2021civ}. We see that the deuteron EDM receives only contributions from the one-pion exchange CP-odd nuclear force, while the nucleon EDM and the contact CP-odd nuclear force (the last term of Eq.~(\ref{fig-LCP})) also contribute to the $^3$He EDM. The coupling constant of the contact CP-odd nuclear force is given by $\bar C_1 = \frac{m_{CP} C_S}{m_N^2} = (3.1\pm 1.6)$ GeV$^{-1}$ which was obtained by chirally rotating the CP-even contact nuclear force, in a similar way as Eq. (\ref{eq:nucleonEDM}) \cite{Yamanaka:2021doa}, with $C_S =-120.8$ GeV$^{-2}$ \cite{Epelbaum:2008ga}.
    The coefficient of $\bar C_1$ was calculated by matching a smeared delta function with the CP-odd $\omega$-exchange nuclear force \cite{Song:2012yh,Yamanaka:2015qfa,Froese:2021civ}, and has a wide uncertainty band, due to the poorly known two-nucleon wavefunction near the origin.
    There is also a calculation within $\chi$EFT which yields a larger coefficient \cite{Bsaisou:2014zwa}, but we do not further examine it here because of its large uncertainty.
    It has also recently been pointed out that the contact CP-odd nuclear force is required to unitarize the CP-odd nucleon-nucleon scattering with one-pion exchange \cite{deVries:2020loy}, but we do not consider this effect in this work.
    We furthermore mention here that the coefficients of $\bar C_1$ are not available for atomic EDMs.
    
    Finally, let us derive an explicit constraint on the magnitude of the Weinberg operator from the presently available experimental data, given by the EDM of $^{199}$Hg. According to our analysis, the EDM of $^{199}$Hg is $d^{\rm Hg} = w \, (\mu = 1 {\rm GeV}) \left( - 1.3 \pm 0.96 \right) \times 10^{-2}  e \, {\rm MeV}$ where the central value is given only by the contribution of the intrinsic neutron EDM and the error is obtained by the taking the quadrature of the theoretical uncertainty of the Weinberg operator contribution of the neutron EDM ($60 \%$) \cite{Yamanaka:2020kjo}, the error associated with the nuclear level calculation ($30 \%$) \cite{Yanase:2020oos}, and the maximal value of the our Weinberg operator result (the second term of (\ref{eq-mercury})), which we consider here to be a systematic error. Combined with the experimental result, $\left|d^{\rm Hg \, (exp)}\right| < 7.4 \times 10^{-30} e \, {\rm cm} $ \cite{Graner:2016ses}, this leads to
    \begin{equation}
        \left|w \, ( \mu = 1 {\rm TeV}) \right| < 4 \times 10^{-10} {\rm GeV}^{-2}.
    \end{equation}
    Here, we divided the experimental value by $ (-1.3 + 0.96) \times 10^{-2} e \, {\rm MeV}$, so as to obtain an upper limit. Furthermore, $w \, (\mu = 1 {\rm TeV})= w \, (\mu = 1 {\rm GeV})/0.3$ is renormalized at $1$ TeV \cite{Braaten:1990gq,Braaten:1990zt,Degrassi:2005zd,Dekens:2013zca,deVries:2019nsu,Kley:2021yhn}.
    It is interesting to compare this limit with that given by the direct measurement of the neutron EDM.
    The current experimental constraint $|d_n| < 1.8 \times 10^{-26} e$ cm \cite{Abel:2020pzs} yield an upper limit of
    \begin{equation}
            \left|w \, ( \mu = 1 {\rm TeV}) \right| < 4 \times 10^{-10} {\rm GeV}^{-2},
    \end{equation}
    where we took into account the uncertainty band of Eq. (\ref{eq:nucleonEDM}).
    We see that the EDM of the $^{199}$Hg atom and that of the neutron provide comparable constraints on the Weinberg operator, if we assume that it is the sole source of CP violation at the scale $\mu =1$ TeV.

\section{Conclusions}
\label{Conclusions}
    In this paper we calculated the pion-vacuum transition matrix element of the Weinberg operator using QCD sum rules and then the isovector CP-odd pion-nucleon interaction within $\chi$EFT at NLO. We finally derived relations between the Weinberg operator and the EDMs of several atoms and nuclei which are of experimental interest. For the case of atoms, the contribution of the nucleon-nucleon interaction was found to be of the same order as that induced by the intrinsic nucleon EDM.
    
    Possible improvements of this work include a more accurate investigation of the OPE convergence by taking into account condensates of higher dimension. Furthermore, using the correlator of two Weinberg operators, it was not possible to determine the sign of the matrix element studied in this work. Therefore, it will in the future become necessary to study the correlator between a Weinberg operator and a pion interpolating field.
    
    Nevertheless, our results suggest that, depending on the sign, the CP violation may be enhanced in the $^{129}$Xe EDM through the new Weinberg operator contribution. This experimental observable might therefore become crucial in the future determination of the magnitude of the Weinberg operator. 
\section*{Acknowledgments}
    We thank Junji Hisano and Makoto Oka for fruitful discussions and useful comments. P.G. is supported by the Grant-in-Aid for Scientific Research (C) (JSPS KAKENHI Grant Number JP20K03940) and the Leading Initiative for Excellent Young Researchers (LEADER) of the Japan Society for the Promotion of Science (JSPS).
\appendix

\section{Borel transforms}
\label{Borel transforms}

    We define the Borel transform of a function $f(Q^2)$ with $Q^2=-q^2$ as
    \begin{equation}
        \mathcal{B} \left[ f(Q^2) \right] := \lim_{\substack{Q^2,n \to \infty \\ Q^2/n = M^2 = const.}} \frac{(Q^2)^n}{(n-1)!} \left( - \frac{d}{d Q^2} \right)^n f(Q^2),
    \end{equation}
    where the unphysical parameter $M$ is called the Borel mass. Below, we provide specific expressions for the Borel transforms used in this work.
    \begin{align}
        & \mathcal{B} \left[ \left( \frac{1}{s+Q^2} \right)^k \right] = \frac{1}{(k-1)!} \left( \frac{1}{M^2} \right)^k e^{-s/M^2}, \\
        & \mathcal{B} \left[ (Q^2)^k \log Q^2 \right] = (-1)^{k+1} \Gamma (k+1) (M^2)^k.
    \end{align}
    Here, $k \in \mathbb{N}_+$ and $\Gamma(z)$ denotes the Euler gamma function.

\section{Fourier transforms}
\label{Fourier transforms}

    In this paper, we need to consider Fourier transforms from momentum space to configuration space and vice versa, which will 
    in turn be discussed in this section, mostly following Appendix E of Ref.\cite{Haisch:2019bml}.
    We define the Fourier transform of a function $f(p)$ to configuration space as
    \begin{equation}
        \mathcal{F} \left[ f(p) \right] = \left( \frac{\mu_{\text{IR}}^2}{4 \pi e^{\gamma_E}} \right)^{- \epsilon_{\text{IR}}} \int \frac{d^{4 + 2 \epsilon_{\text{IR}}} p}{(2 \pi)^4} e^{- i p \cdot x} f(p).
        \label{al-fourier transformation}
    \end{equation}
    where the divergent integral is treated using dimensional regularization with $d=4 + 2\epsilon_{\text{IR}}$.
    The reason for adopting this definition is to ensure the conservation of gauge invariance throughout the calculation. 
    The OPE is based on the division of scales, in which higher regions of momentum space can be treated perturbatively 
    to compute the Wilson coefficients \cite{Novikov:1983jt}. Dimensional regularization provides the most economical renormalization scheme 
    to do this in a gauge invariant manner. Any IR-pole can hence be removed as it should be considered to be part of 
    the low momentum space regions and therefore can be absorbed into the renormalization of the condensates.

    For functions of the form $1/(p^2)^k$ and $p^\mu p^\nu/(p^2)^k$, a simple calculation leads to
    \begin{align}
        &\mathcal{F} \left[ \frac{1}{(p^2)^k} \right] = \frac{i}{4^k \pi^2} \left( - \frac{\mu^2_{\text{IR}} x^2}{4 e^{\gamma_E}} \right)^{-\epsilon_{\text{IR}}} \frac{\Gamma(2-k+\epsilon_{\text{IR}})}{\Gamma(k)} (x^2)^{k-2}, \label{al-ft formula1} \\
        &\mathcal{F} \left[ \frac{p_\mu p_\nu}{(p^2)^k} \right] = (i \partial_\mu) (i \partial_\nu) \mathcal{F} \left[ \frac{1}{(p^2)^k} \right], \label{al-ft formula2}
    \end{align}
    where $k\in \mathbb{N}_+$. The Fourier transforms of type (\ref{al-ft formula1}) and (\ref{al-ft formula2}) relevant for our article are
    \begin{align}
        &\mathcal{F} \left[ \frac{1}{p^2} \right] = \frac{i}{4 \pi^2 x^2}, \\
        &\mathcal{F} \left[ \frac{1}{(p^2)^2} \right] = \frac{i}{16 \pi^2} \left[ \frac{1}{\epsilon_{\text{IR}}} - \ln \left( - \frac{\mu^2_{\text{IR}} x^2}{4} \right) \right], \\
        &\mathcal{F} \left[ \frac{1}{(p^2)^3} \right] = - \frac{i}{2 \cdot 4^3 \pi^2} x^2 \left[ \frac{1}{\epsilon_{\text{IR}}} + 1 - \ln \left( - \frac{\mu_{\text{IR}}^2 x^2}{4} \right) \right], \\
        &\mathcal{F} \left[ \frac{1}{(p^2)^4} \right] = \frac{i}{2^{10} \cdot 3 \pi^2} (x^2)^2 \left[ \frac{1}{\epsilon_{\text{IR}}} + \frac{3}{2} - \ln \left(- \frac{\mu^2_{\text{IR}} x^2}{4} \right) \right].
    \end{align}

    In QCD sum rules, one often encounters ultraviolet divergent Fourier transforms of functions $f(x)$ from configuration space to momentum space. In this work, we define them as
    \begin{equation}
        \mathcal{F} \left[ f(x) \right] = \left( \frac{\mu^2_{\text{UV}} e^{\gamma_E}}{4 \pi} \right)^{- \epsilon_{\text{UV}}} \int d^{4-2\epsilon_{\text{UV}}} e^{i p \cdot x} f(x), 
    \end{equation}
    and use again dimensional regularization in treating the UV divergences. For the functions $1/(x^2)^k$ and $x^\mu x^\nu/(x^2)^k$ with $k \in \mathbb{N}_+$,
    we have
    \begin{align}
        &\mathcal{F} \left[ \frac{1}{(x^2)^k} \right] = - \frac{i \pi^2}{4^{k-2}} \left( - \frac{\mu^2_{\text{UV}} e^{\gamma_E}}{p^2} \right)^{- \epsilon_{\text{UV}}} \frac{\Gamma (2-k-\epsilon_{\text{UV}})}{\Gamma (k)} (p^2)^{k-2}, \\
        &\mathcal{F} \left[ \frac{x^\mu x^\nu}{(x^2)^k} \right] = \left( \frac{1}{i} \frac{\partial}{\partial p_\mu} \right)  \left( \frac{1}{i} \frac{\partial}{\partial p_\nu} \right) \mathcal{F} \left[ \frac{1}{(x^2)^k} \right].
    \end{align}
    The resultant expressions will in the calculations of this paper be either Borel transformed or its imaginary part taken. 
    As any polynomial of $p^2$ will vanish through the Borel transform, and the imaginary part of the correlation function originates only from logarithms, 
    we only need to consider the $\log$ terms and can drop all polynomials in all practical calculations of this paper. The following simplified formulas 
    will therefore be sufficient:
    \begin{align}
        &\mathcal{F}_{\text{pra}} \left[ \frac{1}{(x^2)^4} \right] = \frac{\pi^2 i}{2^6 \cdot 3} (-p^2)^2 \ln \left( - p^2 \right), \\
        &\mathcal{F}_{\text{pra}} \left[ \frac{1}{(x^2)^5} \right] = \frac{\pi^2 i}{2^{10} \cdot 3^2} (-p^2)^3 \ln \left( -p^2 \right).
    \end{align}

\section{Details of the OPE calculation}
\label{Details of OPE calculation}
    The correlation function of Eq.~(\ref{al-OPE}) can be expressed using the Weinberg operator of Eq.~(\ref{al-Weinberg operator2}) as
    \begin{equation}
        \begin{split}
            \Pi_{\text{OPE}} (q) &= i \int d^4 x e^{i q \cdot x} \cdot \frac{1}{9} w^2 f^{abc} f^{a'b'c'} \epsilon^{\nu \rho \alpha \beta} \epsilon^{\nu ' \rho ' \alpha ' \beta '} g^{\mu \sigma} g^{\mu ' \sigma '} \\
            & \quad \times \langle 0|T \left[ 2 \left( \partial_\mu A_\nu^a \right) \left( \partial_\rho A_\sigma^c \right) - \left( \partial_\mu A_\nu^a \right) \left( \partial_\sigma A_\rho^c \right) - \left( \partial_\nu A_\mu^a \right) \left( \partial_\rho A_\sigma^c \right) \right] \left( \partial_\alpha A_\beta^b \right) (x) \\
            & \quad \cdot \left[ 2 \left( \partial_{\mu'} A_{\nu'}^{a'} \right) \left( \partial_{\rho'} A_{\sigma'}^{c'} \right) - \left( \partial_{\mu'} A_{\nu'}^{a'} \right) \left( \partial_{\sigma'} A_{\rho'}^{c'} \right) - \left( \partial_{\nu'} A_{\mu'}^{a'} \right) \left( \partial_{\rho'} A_{\sigma'}^{c'} \right) \right] \left( \partial_{\alpha'} A_{\beta'}^{b'} \right) (0) |0 \rangle \\
            & = i w^2 \int d^4 x e^{i q \cdot x} \frac{1}{9} f^{abc} f^{a'b'c'} \\
            & \quad \times \left[ 4 \epsilon^{\nu \rho \alpha \beta} \epsilon^{\nu ' \rho ' \alpha ' \beta '} g^{\mu \sigma} g^{\mu ' \sigma '} -2 \epsilon^{\nu \sigma \alpha \beta} \epsilon^{\nu ' \rho ' \alpha ' \beta '} g^{\mu \rho} g^{\mu ' \sigma '} -2 \epsilon^{\nu \rho \alpha \beta} \epsilon^{\mu ' \rho ' \alpha ' \beta '} g^{\mu \sigma} g^{\nu ' \sigma '} \right. \\
            & \left. \quad -2 \epsilon^{\nu \rho \alpha \beta} \epsilon^{\nu ' \sigma ' \alpha ' \beta '} g^{\mu \sigma} g^{\mu ' \rho '} -2 \epsilon^{\mu \rho \alpha \beta} \epsilon^{\nu ' \rho ' \alpha ' \beta '} g^{\nu \sigma} g^{\mu ' \sigma '} + \epsilon^{\nu \sigma \alpha \beta} \epsilon^{\nu ' \sigma ' \alpha ' \beta '} g^{\mu \rho} g^{\mu ' \rho '} \right. \\
            & \quad \left. + \epsilon^{\nu \sigma \alpha \beta} \epsilon^{\mu ' \rho ' \alpha ' \beta '} g^{\mu \rho} g^{\nu ' \sigma '} + \epsilon^{\mu \rho \alpha \beta} \epsilon^{\nu ' \sigma ' \alpha ' \beta '} g^{\nu \sigma} g^{\mu ' \rho '} + \epsilon^{\mu \rho \alpha \beta} \epsilon^{\mu ' \rho ' \alpha ' \beta '} g^{\nu \sigma} g^{\nu ' \sigma '} \right] \\
            & \quad \times \Braket{0|T \left[ (\partial_\mu A^a_{\nu} ) (\partial_\alpha A^b_\beta) (\partial_\rho A^c_\sigma) (x) (\partial_{\mu '} A^{a'}_{\nu '} ) (\partial_{\alpha '} A^{b'}_{\beta '}) (\partial_{\rho '} A^{c'}_{\sigma '}) (0) \right] |0},
        \end{split}
        \label{al-OPE formula1}
    \end{equation}
    where, we have omitted all higher order contributions with respect to the strong coupling constant. The Wick contraction of the matrix element given in the last line of Eq.~(\ref{al-OPE formula1}) yields six distinct terms, which can be related to each other by simple permutations. It is hence sufficient to evaluate only one term, which we define as
    \begin{equation}
        \Pi^{abc;a'b'c'}_{\mu \nu \alpha \beta \rho \sigma ; \mu ' \nu ' \alpha ' \beta ' \rho ' \sigma '} (x) = \left. \left( \partial_\mu^{(x)} \partial_{\mu '}^{(y)} S^{a a'}_{\nu \nu '} (x-y) \right) \left( \partial_\alpha^{(x)} \partial_{\alpha '}^{(y)} S^{b b'}_{\beta \beta '} (x-y) \right) \left( \partial_\rho^{(x)} \partial_{\rho '}^{(y)} S^{c c'}_{\sigma \sigma '} (x-y) \right) \right|_{y=0}, 
        \label{al-OPE formula3}
    \end{equation}
    where $S^{ab}_{\mu \nu}(x-y)$ stands for the gluon propagator. 
    Other terms generated by the Wick contraction can be obtained by permutating the 
    gauge and Lorentz indices in $\Pi^{abc;a'b'c'}_{\mu \nu \alpha \beta \rho \sigma; \mu ' \nu ' \alpha ' \beta ' \rho ' \sigma '}(x)$. Gauge indices can be factored out as $\delta^{ab}$ from each gluon propagator. Eq.~(\ref{al-OPE formula1}) can therefore be simplified as,
    \begin{equation}
        \begin{split}
            &\Pi_{\text{OPE}} (q) \\
            & \quad 
            \begin{split}
                &= i w^2 \int d^4 x e^{i q \cdot x} \frac{1}{9} f^{abc} f^{a'b'c'} \\
                & \quad \times \left[ 4 \epsilon^{\nu \rho \alpha \beta} \epsilon^{\nu ' \rho ' \alpha ' \beta '} g^{\mu \sigma} g^{\mu ' \sigma '} -2 \epsilon^{\nu \sigma \alpha \beta} \epsilon^{\nu ' \rho ' \alpha ' \beta '} g^{\mu \rho} g^{\mu ' \sigma '} -2 \epsilon^{\nu \rho \alpha \beta} \epsilon^{\mu ' \rho ' \alpha ' \beta '} g^{\mu \sigma} g^{\nu ' \sigma '} \right. \\
                & \left. \quad \quad -2 \epsilon^{\nu \rho \alpha \beta} \epsilon^{\nu ' \sigma ' \alpha ' \beta '} g^{\mu \sigma} g^{\mu ' \rho '} -2 \epsilon^{\mu \rho \alpha \beta} \epsilon^{\nu ' \rho ' \alpha ' \beta '} g^{\nu \sigma} g^{\mu ' \sigma '} + \epsilon^{\nu \sigma \alpha \beta} \epsilon^{\nu ' \sigma ' \alpha ' \beta '} g^{\mu \rho} g^{\mu ' \rho '} \right. \\
                & \quad \quad \left. + \epsilon^{\nu \sigma \alpha \beta} \epsilon^{\mu ' \rho ' \alpha ' \beta '} g^{\mu \rho} g^{\nu ' \sigma '} + \epsilon^{\mu \rho \alpha \beta} \epsilon^{\nu ' \sigma ' \alpha ' \beta '} g^{\nu \sigma} g^{\mu ' \rho '} + \epsilon^{\mu \rho \alpha \beta} \epsilon^{\mu ' \rho ' \alpha ' \beta '} g^{\nu \sigma} g^{\nu ' \sigma '} \right] \\
                & \quad \times \left[ \delta^{a a'} \delta^{bb'} \delta^{cc'} \Pi_{\mu \nu \alpha \beta \rho \sigma ; \mu ' \nu ' \alpha ' \beta ' \rho ' \sigma '} (x) + \delta^{a a'} \delta^{bc'} \delta^{cb'} \Pi_{\mu \nu \alpha \beta \rho \sigma ; \mu ' \nu ' \rho ' \sigma ' \alpha ' \beta '} (x) \right. \\
                & \quad \left. \quad  + \delta^{a b'} \delta^{ba'} \delta^{cc'} \Pi_{\mu \nu \alpha \beta \rho \sigma ; \alpha ' \beta ' \mu ' \nu ' \rho ' \sigma '} (x) + \delta^{a b'} \delta^{bc'} \delta^{ca'} \Pi_{\mu \nu \alpha \beta \rho \sigma ; \alpha ' \beta ' \rho ' \sigma ' \mu ' \nu '} (x) \right. \\
                & \quad \left. \quad + \delta^{a c'} \delta^{bb'} \delta^{ca'} \Pi_{\mu \nu \alpha \beta \rho \sigma ; \rho ' \sigma ' \alpha ' \beta ' \mu ' \nu '} (x) + \delta^{a c'} \delta^{ba'} \delta^{cb'} \Pi_{\mu \nu \alpha \beta \rho \sigma ; \rho ' \sigma ' \mu ' \nu ' \alpha ' \beta '} (x) \right] \\
            \end{split} \\
            & \quad \begin{split}
                = i w^2 \frac{2^3}{3} \int d^4 x e^{i q \cdot x} C^{\mu \nu \alpha \beta \rho \sigma ; \mu ' \nu ' \alpha ' \beta ' \rho ' \sigma '} & \left[ \Pi_{\mu \nu \alpha \beta \rho \sigma ; \mu ' \nu ' \alpha ' \beta ' \rho ' \sigma '} (x) - \Pi_{\mu \nu \alpha \beta \rho \sigma ; \mu ' \nu ' \rho ' \sigma ' \alpha ' \beta '} (x) \right. \\
                & \left. - \Pi_{\mu \nu \alpha \beta \rho \sigma ; \alpha ' \beta ' \mu ' \nu ' \rho ' \sigma '} (x) + \Pi_{\mu \nu \alpha \beta \rho \sigma ; \alpha ' \beta ' \rho ' \sigma ' \mu ' \nu '} (x) \right. \\
                & \left. - \Pi_{\mu \nu \alpha \beta \rho \sigma ; \rho ' \sigma ' \alpha ' \beta ' \mu ' \nu '} (x) + \Pi_{\mu \nu \alpha \beta \rho \sigma ; \rho ' \sigma ' \mu ' \nu ' \alpha ' \beta '} (x) \right],
              \end{split}
        \end{split}
        \label{al-OPE master formula1}
    \end{equation}
    where $\Pi_{\mu \nu \alpha \beta \rho \sigma ; \mu ' \nu ' \alpha ' \beta ' \rho ' \sigma '} (x)$ is defined as 
    \begin{equation}
    \Pi^{abc;a'b'c'}_{\mu \nu \alpha \beta \rho \sigma ; \mu ' \nu ' \alpha ' \beta ' \rho ' \sigma '} (x) = \delta^{a a'} \delta^{bb'} \delta^{cc'} 
    \Pi_{\mu \nu \alpha \beta \rho \sigma ; \mu ' \nu ' \alpha ' \beta ' \rho ' \sigma '} (x),
    \end{equation}
    and
    \begin{equation}
        \begin{split}
            C^{\mu \nu \alpha \beta \rho \sigma ; \mu ' \nu ' \alpha ' \beta ' \rho ' \sigma '} := & \left[ 4 \epsilon^{\nu \rho \alpha \beta} \epsilon^{\nu ' \rho ' \alpha ' \beta '} g^{\mu \sigma} g^{\mu ' \sigma '} -2 \epsilon^{\nu \sigma \alpha \beta} \epsilon^{\nu ' \rho ' \alpha ' \beta '} g^{\mu \rho} g^{\mu ' \sigma '} -2 \epsilon^{\nu \rho \alpha \beta} \epsilon^{\mu ' \rho ' \alpha ' \beta '} g^{\mu \sigma} g^{\nu ' \sigma '} \right. \\
            & \left. -2 \epsilon^{\nu \rho \alpha \beta} \epsilon^{\nu ' \sigma ' \alpha ' \beta '} g^{\mu \sigma} g^{\mu ' \rho '} -2 \epsilon^{\mu \rho \alpha \beta} \epsilon^{\nu ' \rho ' \alpha ' \beta '} g^{\nu \sigma} g^{\mu ' \sigma '} + \epsilon^{\nu \sigma \alpha \beta} \epsilon^{\nu ' \sigma ' \alpha ' \beta '} g^{\mu \rho} g^{\mu ' \rho '} \right. \\
            & \left. + \epsilon^{\nu \sigma \alpha \beta} \epsilon^{\mu ' \rho ' \alpha ' \beta '} g^{\mu \rho} g^{\nu ' \sigma '} + \epsilon^{\mu \rho \alpha \beta} \epsilon^{\nu ' \sigma ' \alpha ' \beta '} g^{\nu \sigma} g^{\mu ' \rho '} + \epsilon^{\mu \rho \alpha \beta} \epsilon^{\mu ' \rho ' \alpha ' \beta '} g^{\nu \sigma} g^{\nu ' \sigma '} \right].
          \end{split}
      \label{al-OPE master formula2}
    \end{equation}

    For convenience of the later calculations, we expand the gluon propagator in terms of condensates and other perturbative terms relevant for 
    this work. As we are here especially interested in the quark-mass dependent part of the correlator, it is necessary to 
    take into account the leading order quark loop insertion of the gluon propagator. 
    The gluon propagator can thus be expanded as
    \begin{equation}
        S^{ab}_{\mu \nu} (x,y) = \left( S^{(0)} \right)^{ab}_{\mu \nu} (x-y) + \left( S^{(\text{loop})} \right)^{ab}_{\mu \nu} (x-y) + \left( S^{(G)} \right)^{ab}_{\mu \nu} (x,y) + \left( S^{(q)} \right)^{ab}_{\mu \nu} (x-y),
    \end{equation}
    where $S^{(0)}(x-y)$ is the free gluon propagator, $S^{(\text{loop})}(x-y)$ its leading order perturbative quark loop correction, $S^{(G)}(x,y)$ stands for the gluon condensate insertion and $S^{(q)}(x-y)$ for the quark loop correction in which one of the quark propagator is replaced by the quark condensate. 
    The analytic form of these terms can be given as 
    \begin{align}
        & \left( S^{(0)} \right)^{ab}_{\mu \nu} (x-y) = \frac{\delta^{ab} g_{\mu \nu}}{4 \pi^2 (x-y)^2}, \\
        & \left( S^{(\text{loop})} \right)^{ab}_{\mu \nu} (x-y) = \frac{m_q^2 \alpha_s}{4^3 \pi^2} \delta^{ab} \left[ 3 g_{\mu \nu} \ln \left( - \frac{\mu_{\text{IR}}^2 (x-y)^2}{4} \right) - 2 \frac{(x-y)_\mu (x-y)_\nu}{(x-y)^2} \right], \\
        & \left( S^{(q)} \right)^{ab}_{\mu \nu} (x-y) = \frac{m_q \alpha_s}{2^{6} \cdot 3^2 \pi} \Braket{\bar{q}q} \delta^{ab} \left[ 5 g^{\mu \nu} (x-y)^2 + \left( -5 g^{\mu \nu} (x-y)^2 + 2 (x-y)^\mu (x-y)^\nu \right) \ln \left( - \frac{\mu_{\text{IR}}^2 (x-y)^2}{4} \right) \right], \\
        & \left( S^{(G)} \right)^{ab}_{\mu \nu} (x,y) = x^{\mu '} y^{\nu '} \frac{\Braket{G^2}}{2^7 \cdot 3} \delta^{ab} (g_{\mu ' \nu '} g_{\mu \nu} - g_{\mu ' \nu} g_{\mu \nu '}), \\
        & \left( S^{(q)} \right)^{ab}_{\mu \nu} (x-y) = \frac{m_q \alpha_s}{2^{6} \cdot 3^2 \pi} \Braket{\bar{q}q} \delta^{ab} \left[ 5 g^{\mu \nu} (x-y)^2 + \left( -5 g^{\mu \nu} (x-y)^2 + 2 (x-y)^\mu (x-y)^\nu \right) \ln \left( - \frac{\mu_{\text{IR}}^2 (x-y)^2}{4} \right) \right].
    \end{align}
    For obtaining $S^{(\text{loop})}(x-y)$ we have expanded the full expression around the 
    zero quark mass limit and only kept the second order $m_q^2$ term, since the leading order term does not contribute to the quark mass dependence 
    and the first order term vanishes. Similarly, for $S^{(q)}(x-y)$, the leading order $m_q^0$ term vanishes and we keep 
    only the first non-vanishing $m_q$ term. 

    Let us here, as an example, show the explicit further steps needed to calculate the OPE term corresponding to 
    the left-most the diagram of Fig.~\ref{fig-diagrams}. It can be written down as
    \begin{equation}
        \begin{split}
            &(\Pi_{\text{loop}}^{\prime})^{abc;a'b'c'}_{\mu \nu \alpha \beta \rho \sigma ; \mu ' \nu ' \alpha ' \beta ' \rho ' \sigma '} (x) \\
            &= \left. \left( \partial^{(x)}_\mu \partial^{(y)}_{\mu '} (S^{(\text{loop})})^{aa'}_{\nu \nu '} (x-y) \right) \cdot \left( \partial^{(x)}_\alpha \partial^{(y)}_{\alpha '} (S^{(0)})^{bb'}_{\beta \beta '} (x-y) \right) \cdot \left( \partial^{(x)}_\rho \partial^{(y)}_{\rho '} (S^{(0)})^{cc'}_{\sigma \sigma '} (x-y) \right) \right|_{y=0} \\
            & \left. + \left( \partial^{(x)}_\mu \partial^{(y)}_{\mu '} (S^{(0)})^{aa'}_{\nu \nu '} (x-y) \right) \cdot \left( \partial^{(x)}_\alpha \partial^{(y)}_{\alpha '} (S^{(\text{loop})})^{bb'}_{\beta \beta '} (x-y) \right) \cdot \left( \partial^{(x)}_\rho \partial^{(y)}_{\rho '} (S^{(0)})^{cc'}_{\sigma \sigma '} (x-y) \right) \right|_{y=0} \\
            & \left. + \left( \partial^{(x)}_\mu \partial^{(y)}_{\mu '} (S^{(0)})^{aa'}_{\nu \nu '} (x-y) \right) \cdot \left( \partial^{(x)}_\alpha \partial^{(y)}_{\alpha '} (S^{(0)})^{bb'}_{\beta \beta '} (x-y) \right) \cdot \left( \partial^{(x)}_\rho \partial^{(y)}_{\rho '} (S^{(\text{loop})})^{cc'}_{\sigma \sigma '} (x-y) \right) \right|_{y=0} \\
            & = \delta^{aa'} \delta^{bb'} \delta^{cc'} (\Pi_{\text{loop}})_{\mu \nu \alpha \beta \rho \sigma ; \mu ' \nu ' \alpha ' \beta ' \rho ' \sigma '} (x),
        \end{split}
    \end{equation}
    where
    \begin{equation}
        \begin{split}
        &(\Pi_{\text{loop}}^{\prime})_{\mu \nu \alpha \beta \rho \sigma ; \mu ' \nu ' \alpha ' \beta ' \rho ' \sigma '} (x) = \sum_{u,d} \frac{m_q^2 \alpha_s}{2^7 \pi^6} \\
        & \quad
        \begin{split}
            \times & \left\{ g_{\beta \beta '} g_{\sigma \sigma '} \left( \frac{g_{\alpha \alpha '}}{x^4} - 4 \frac{x_\alpha x_{\alpha '}}{x^6} \right) \left( \frac{g_{\rho \rho '}}{x^4} - 4 \frac{x_\rho x_{\rho '}}{x^6} \right) \left[(3 g_{\mu \mu '} g_{\nu \nu '} - g_{\mu \nu} g_{\mu ' \nu '} - g_{\mu \nu '} g_{\mu ' \nu}) \frac{1}{x^2} \right. \right. \\
            & \left. \left. + 2 (g_{\mu \nu} x_{\mu '} x_{\nu '} + g_{\mu \nu '} x_{\mu '} x_\nu + g_{\mu ' \nu} x_\mu x_{\nu '} + g_{\mu ' \nu '} x_\mu x_\nu + g_{\mu \mu '} x_\nu x_{\nu '} - 3 g_{\nu \nu '} x_\mu x_{\mu '}) \frac{1}{x^4} - 8 \frac{x_\mu x_{\mu '} x_\nu x_{\nu'}}{x^6} \right] \right. \\
            & \left. + g_{\nu \nu '} g_{\sigma \sigma '} \left( \frac{g_{\mu \mu '}}{x^4} - 4 \frac{x_\mu x_{\mu '}}{x^6} \right) \left( \frac{g_{\rho \rho '}}{x^4} - 4 \frac{x_\rho x_{\rho '}}{x^6} \right) \left[(3 g_{\alpha \alpha '} g_{\beta \beta '} - g_{\alpha \beta} g_{\alpha ' \beta '} - g_{\alpha \beta '} g_{\alpha ' \beta}) \frac{1}{x^2} \right. \right. \\
            & \left. \left. + 2 (g_{\alpha \beta} x_{\alpha '} x_{\beta '} + g_{\alpha \beta '} x_{\alpha '} x_\beta + g_{\alpha ' \beta} x_\alpha x_{\beta '} + g_{\alpha ' \beta '} x_\alpha x_\beta + g_{\alpha \alpha '} x_\beta x_{\beta '} - 3 g_{\beta \beta '} x_\alpha x_{\alpha '}) \frac{1}{x^4} - 8 \frac{x_\alpha x_{\alpha '} x_\beta x_{\beta'}}{x^6} \right] \right. \\
            & \left. + g_{\beta \beta '} g_{\nu \nu '} \left( \frac{g_{\alpha \alpha '}}{x^4} - 4 \frac{x_\alpha x_{\alpha '}}{x^6} \right) \left( \frac{g_{\mu \mu '}}{x^4} - 4 \frac{x_\mu x_{\mu '}}{x^6} \right) \left[(3 g_{\rho \rho '} g_{\sigma \sigma '} - g_{\rho \sigma} g_{\rho ' \sigma '} - g_{\rho \sigma '} g_{\rho ' \sigma}) \frac{1}{x^2} \right. \right. \\
            & \left. \left. + 2 (g_{\rho \sigma} x_{\rho '} x_{\sigma '} + g_{\rho \sigma '} x_{\rho '} x_\sigma + g_{\rho ' \sigma} x_\rho x_{\sigma '} + g_{\rho ' \sigma '} x_\rho x_\sigma + g_{\rho \rho '} x_\sigma x_{\sigma '} - 3 g_{\sigma \sigma '} x_\rho x_{\rho '}) \frac{1}{x^4} - 8 \frac{x_\rho x_{\rho '} x_\sigma x_{\sigma'}}{x^6} \right] \right\}.
        \end{split}
        \end{split}
        \label{al-cor loop}
    \end{equation}
    Using Eqs.~(\ref{al-OPE master formula2}) and (\ref{al-cor loop}), $\Pi^{\prime}_{\text{loop}} (q^2)$, which denotes the contribution of these quark loop diagrams to Eq.~(\ref{al-OPE master formula1}), can then be computed as
    \begin{equation}
        \begin{split}
            &\Pi_{\text{loop}}^{\prime} (q^2) \\
            & \quad \begin{split}
                = i w^2 \frac{2^3}{3} \int d^4 x e^{i q \cdot x} C^{\mu \nu \alpha \beta \rho \sigma ; \mu ' \nu ' \alpha ' \beta ' \rho ' \sigma '} & \left[ (\Pi_{\text{loop}}^{\prime})_{\mu \nu \alpha \beta \rho \sigma ; \mu ' \nu ' \alpha ' \beta ' \rho ' \sigma '} (x) - (\Pi_{\text{loop}}^{\prime})_{\mu \nu \alpha \beta \rho \sigma ; \mu ' \nu ' \rho ' \sigma ' \alpha ' \beta '} (x) \right. \\
                & \left. - (\Pi_{\text{loop}}^{\prime})_{\mu \nu \alpha \beta \rho \sigma ; \alpha ' \beta ' \mu ' \nu ' \rho ' \sigma '} (x) + (\Pi_{\text{loop}}^{\prime})_{\mu \nu \alpha \beta \rho \sigma ; \alpha ' \beta ' \rho ' \sigma ' \mu ' \nu '} (x) \right. \\
                & \left. -(\Pi_{\text{loop}}^{\prime})_{\mu \nu \alpha \beta \rho \sigma ; \rho ' \sigma ' \alpha ' \beta ' \mu ' \nu '} (x) + (\Pi_{\text{loop}}^{\prime})_{\mu \nu \alpha \beta \rho \sigma ; \rho ' \sigma ' \mu ' \nu ' \alpha ' \beta '} (x) \right]
            \end{split} \\
            & = i w^2 \frac{m_q^2 \alpha_s}{2^4 \cdot 3 \pi^6} \int d^4 x e^{i q \cdot x} \frac{6912}{x^{10}} \\
            & = - w^2 \frac{m_q^2 \alpha_s}{2^{6} \pi^4} (-q^2)^3 \ln \left( -q^2 \right).
        \end{split}
    \end{equation}
    Taking into account the contributions of both u and d quarks, noting that 
    \begin{equation}
        \begin{split}
            m_u^2 + m_d^2 &= \frac{1}{4} \left[ (m_+ + m_-)^2 + (m_+ - m_-)^2 \right]  \\
            &= \frac{1}{2} (m_+^2 + m_-^2), 
        \end{split}
    \end{equation}
    where $m_+ \equiv m_u + m_d$ and $m_- \equiv m_u - m_d$, and retaining only the $m_-$-dependent part, we have 
    \begin{equation}
        \Pi_{\text{loop}} (q^2, m_- \neq 0) = - w^2 \frac{ m_-^2 \alpha_s}{2^7 \pi^4} (-q^2)^3 \ln \left( -q^2 \right).
    \end{equation}

    For the term involving the quark condensate, $\Pi^{(q)}(q^2)$, chirality demands that it must be proportional to $\sum_{u,d} m_q \Braket{\bar{q}q}$ 
    at leading non-vanishing order in the quark mass expansion. The $m_-$-dependence can be extracted from this expression as 
        \begin{align}
            &
            \begin{aligned}
              \sum_{q=u,d} m_q \Braket{\bar{q} q} &= m_u \Braket{\bar{u} u} + m_d \Braket{\bar{d} d}  \\
              & = m_u (\Braket{\bar{q} q} - 2 m_- B_0^2 h_3) + m_d (\Braket{\bar{q} q} + 2 m_- B_0^2 h_3)  \\
              & = m_+ \Braket{\bar{q} q} - 2 m^2_- B_0^2 h_3,
            \end{aligned}
            \label{eq-qmassdependense}
        \end{align}
        where
        \begin{equation}
        \begin{split}
            \Braket{\bar{q}q} &\equiv (\langle\bar{u}u\rangle+\langle\bar{d}d \rangle)/2, \\
            \Braket{\bar{u} u - \bar{d} d} &= -4 (m_u - m_d) B_0^2 h_3.
        \end{split}
        \label{eq-parameters}
        \end{equation}
        Numerically, we will use the values provided in Ref.~\cite{GomezNicola:2010vgd} (see Table.~\ref{cond_value}).

\section{One-loop correction to $\bar g_{\pi NN}^{(1)}$\label{eq:NLOgpiNN}}

          In this appendix we recapitulate the calculation of the NLO correction to $\bar g_{\pi NN}^{(1)}$ induced by the three-pion interaction of Eq. (\ref{fig-LCP}).
          We assume the following chiral $SU(3)_L \times SU(3)_R$ invariant effective Lagrangian
          \begin{eqnarray}
          {\cal L}_0 &=& \frac{f_\pi^2}{4} {\rm Tr} (\partial_\mu U \partial^\mu U^\dagger ) + {\rm Tr} [\bar B (i\partial \hspace{-.5em}/ - m_B )B] \nonumber\\
          && + \frac{i}{2} {\rm Tr} [\bar B \gamma_\mu (\xi \partial^\mu \xi^\dagger +\xi^\dagger \partial^\mu \xi)B] 
          + \frac{i}{2} {\rm Tr} [\bar B \gamma_\mu B (\partial^\mu \xi \xi^\dagger +\partial^\mu \xi^\dagger \xi)] \nonumber\\
          &&+ \frac{i}{2} (D+F) {\rm Tr} [\bar B \gamma_\mu \gamma_5 (\xi \partial^\mu \xi^\dagger -\xi^\dagger \partial^\mu \xi)B ]
          \nonumber\\
          &&- \frac{i}{2} (D-F) {\rm Tr} [\bar B \gamma_\mu \gamma_5 B (\partial^\mu \xi \xi^\dagger -\partial^\mu \xi^\dagger \xi) ] \ ,
          \label{eq:chiral_lagrangian}
          \end{eqnarray}
          where $D+F = g_A = 1.27$, and
          \begin{equation}
          \xi = \exp \left( \frac{i\phi}{\sqrt{2}f_\pi } \right) \ ,
          \end{equation}
          with $U=\xi^2$.
          The meson field $\phi$ is defined by
          \begin{equation}
          \phi=
          \left(
          \begin{array}{ccc}
          \frac{\pi^0}{\sqrt{2}}+\frac{\eta^0}{\sqrt{6}}&\pi^+&K^+\\
          \pi^- &-\frac{\pi^0}{\sqrt{2}}+\frac{\eta^0}{\sqrt{6}}&K^0 \\
          K^- & \bar K^0 & -2 \frac{\eta^0}{\sqrt{6}}
          \end{array}
          \right) \ ,
          \end{equation}
          and the baryon field $B$ by
          \begin{equation}
          B=
          \left(
          \begin{array}{ccc}
          \frac{\Sigma^0}{\sqrt{2}}+\frac{\Lambda^0}{\sqrt{6}}&\Sigma^+&p\\
          \Sigma^- &-\frac{\Sigma^0}{\sqrt{2}}+\frac{\Lambda^0}{\sqrt{6}}&n \\
          \Xi^- & \Xi^0 & -2 \frac{\Lambda^0}{\sqrt{6}}
          \end{array}
          \right) \ .
          \end{equation}

          We use the heavy baryon approximation where several simplifications occur.
          The 4-momentum of the nucleon can be split into two terms according to the scale separation between the heavy baryon mass and the nonrelativistic spatial 3-momentum, namely $p_\mu = m_N v_\mu + k_\mu$, where $v$ is the velocity 4-vector of the nucleon.
          The baryon propagator is approximated as
          \begin{equation}
          \frac{i}{p\hspace{-0.5em}/\, -m_B +i \epsilon}
          \approx
          \frac{i{\cal P}_+}{v \cdot k + i\epsilon}
          ,
          \end{equation}
          where ${\cal P}_\pm \equiv \frac{1}{2} (1\pm v\hspace{-0.4em}/\,)$. The sign of the causal infinitesimal term ($i\epsilon$) controls the sign of the final result.
          
          The leading order meson-baryon interaction becomes
          \begin{eqnarray}
          {\cal L}_{mB}
          &=&
          \frac{i}{2} (D+F) {\rm Tr} [\bar B \gamma_\mu \gamma_5 (\xi \partial^\mu \xi^\dagger -\xi^\dagger \partial^\mu \xi)B ]
          \nonumber\\
          &= &
          \frac{g_A}{\sqrt{2}f_\pi } {\rm Tr} [\bar B \gamma_\mu \gamma_5 (\partial^\mu \phi ) B ] 
          +\cdots
          \nonumber\\
          &= &
          \frac{g_A}{\sqrt{2}f_\pi } 
          \Biggl[
          \bar p \gamma_\mu \gamma_5 n \partial^\mu \pi^+
          +\bar n \gamma_\mu \gamma_5 p \partial^\mu \pi^-
          +\frac{1}{\sqrt{2}}\bar p \gamma_\mu \gamma_5 p \partial^\mu \pi^0
          -\frac{1}{\sqrt{2}}\bar p \gamma_\mu \gamma_5 p \partial^\mu \pi^0
          \Biggr] 
          +\cdots
          \nonumber\\
          &\approx &
          \frac{\sqrt{2} g_A}{f_\pi } 
          \Biggl[
          \bar H_p S_\mu H_n \partial^\mu \pi^+
          +\bar H_n S_\mu H_p \partial^\mu \pi^-
          +\frac{1}{\sqrt{2}}\bar H_p S_\mu H_p \partial^\mu \pi^0
          -\frac{1}{\sqrt{2}}\bar H_p S_\mu H_p \partial^\mu \pi^0
          \Biggr] 
          +\cdots
          ,
          \end{eqnarray}
          where the ellipses denote irrelevant higher order terms as well as interactions involving strangeness. The four-vector $S^\mu = \frac{1}{2}(0,\vec{\sigma})$ is the Pauli-Lubanski spin vector.

          The NLO correction to $\bar g_{\pi NN}^{(1)}$ induced by the three-pion interaction of Eq. (\ref{fig-LCP}) is then given by
          \begin{eqnarray}
          i{\cal M}_{\pi^\pm}
          &=&
          -2 m_N \Delta_{3\pi}
          \frac{ 2 g_A^2}{ f_\pi^2}
          \int \frac{d^4 k}{(2 \pi )^4}
          \frac{\bar H_N (-i k\cdot S) {\cal P}_+ (ik\cdot S) H_N}{[k^2 - m_\pi^2 ]^2 [ v\cdot k +i\epsilon ]}
          ,
          \\
          i{\cal M}_{\pi^0}
          &=&
          -6 m_N \Delta_{3\pi}
          \frac{ g_A^2}{ f_\pi^2}
          \int \frac{d^4 k}{(2 \pi )^4}
          \frac{\bar H_N (-i k\cdot S) {\cal P}_+ (ik\cdot S) H_N}{[k^2 - m_\pi^2 ]^2 [ v\cdot k +i\epsilon ]}
          ,
          \end{eqnarray}
          where $i{\cal M}_{\pi^\pm}$ and $i{\cal M}_{\pi^0}$ are the contributions from the charged and neutral pion loops, respectively (see Fig.~\ref{fig-gbar NLO}).

          The loop integral can be transformed further, as
          \begin{eqnarray}
          I
          &=&
          \int \frac{d^4 k}{(2 \pi )^4}
          \frac{\bar H_N (S\cdot k) {\cal P}_+ (S\cdot k) H_N}{[k^2 - m_\pi^2 ]^2 [ v\cdot k +i\epsilon ]}
          \nonumber\\
          &=&
          \int \frac{d^4 k}{(2 \pi )^4}
          \frac{\bar H_N (S\cdot k)^2 H_N}{[k^2 - m_\pi^2 
          ]^2 [ v\cdot k +i\epsilon ]}
          \nonumber\\
          &=&
          \int \frac{d^4 k}{(2 \pi )^4}
          \frac{\bar H_N k_\alpha k_\beta \frac{1}{2} \Bigl( [S^\alpha , S^\beta ] +\{S^\alpha , S^\beta \}  \Bigr) H_N}{[k^2 - m_\pi^2 
          ]^2 [ v\cdot k +i\epsilon ]}
          \nonumber\\
          &=&
          \int \frac{d^4 k}{(2 \pi )^4}
          \frac{\bar H_N k_i k_j \frac{1}{4} \delta_{ij} H_N}{[k^2 - m_\pi^2 
          ]^2 [ v\cdot k +i\epsilon ]}
          \nonumber\\
          &=&
          \frac{1}{4}
          \int \frac{d^4 k}{(2 \pi )^4}
          \frac{|{\bf k}|^2 \bar H_N H_N}{[k^2 - m_\pi^2 
          ]^2 [ k_0 +i\epsilon ]}
          \nonumber\\
          &=&
          \frac{1}{4}
          \int \frac{d^4 k}{(2 \pi )^4}
          \frac{|{\bf k}|^2 \bar H_N H_N [({\rm Principal \ value})-i\pi \delta (k_0)]}{[k_0^2-|{\bf k}|^2 - m_\pi^2 
          ]^2 }
          \nonumber\\
          &=&
          \frac{-i}{8}
          \int \frac{d^3 k}{(2 \pi )^3}
          \frac{|{\bf k}|^2 \bar H_N H_N }{[|{\bf k}|^2 + m_\pi^2 
          ]^2 }
          .
          \end{eqnarray}
          We then shift the dimension to use dimensional regularization:
          \begin{eqnarray}
          I
          &=&
          \frac{-i}{8}
          \int \frac{d^d k}{(2 \pi )^d}
          \frac{|{\bf k}|^2 \bar H_N H_N }{[|{\bf k}|^2 + m_\pi^2 
          ]^2 }
          \nonumber\\
          &=&
          \frac{-i}{8}
          \frac{1}{(4 \pi )^{d/2}} \frac{d}{2} \frac{\Gamma (2-\frac{d}{2}-1)}{\Gamma (2)}
          \Biggl( \frac{ 1 }{ m_\pi^2  } \Biggr)^{2-\frac{d}{2}-1}
          \bar H_N H_N
          .
          \end{eqnarray}
          Taking the limit $d\to 3$, we derive
          \begin{eqnarray}
          I
          &=&
          \frac{-i}{8}
          \frac{3}{2(4 \pi )^{3/2}} \Gamma \Biggl(-\frac{1}{2} \Biggr)\,
          m_\pi 
          \bar H_N H_N
          \nonumber\\
          &=&
          i\frac{3  m_\pi}{64 \pi}
          \bar H_N H_N
          ,
          \label{eq:result_ChPT3_pion_loop}
          \end{eqnarray}
          where we used $\Gamma \Bigl(-\frac{1}{2} \Bigr) = -\sqrt{4\pi}$.
          
          The NLO correction to $\bar g_{\pi NN}^{(1)}$ induced by the three-pion interaction is thus obtained as
          \begin{eqnarray}
          \bar g_{\pi NN,{\rm NLO}}^{(1)}
          =
          {\cal M}_{\pi^\pm}
          +
          {\cal M}_{\pi^0}
          =
          -10 m_N \Delta_{3\pi}
          \frac{ g_A^2}{ f_\pi^2}
          \times
          \frac{I}{i}
          =
          -m_N \Delta_{3\pi}
          \frac{15 g_A^2 m_\pi}{32 \pi f_\pi^2}
          ,
          \end{eqnarray}
          which reproduces the formula of Eq.~(\ref{eq:piNNNLO}).

\end{document}